\documentclass[%
onecolumn,
 superscriptaddress,
nofootinbib,
longbibliography,
aps,
prx
]{revtex4-2}

\usepackage{amsmath,amssymb,bm,braket,url,dsfont,xspace}
\usepackage{graphicx}
\usepackage{xcolor}
\usepackage{array}
\usepackage{booktabs}
\usepackage{ulem}

\newcolumntype{L}{>{$}l<{$}} 
\newcolumntype{C}{>{$}c<{$}} 

\usepackage[
draft=false,
bookmarks=true,
bookmarksnumbered=false,
bookmarksopen=false,
bookmarksopenlevel=4,
colorlinks=true,
anchorcolor=black,
citecolor=blue,
filecolor=magenta,
linkcolor=blue,
linkbordercolor={1 0 0},
urlcolor=blue,
pdfborder={0 0 1},
pdftitle={},
pdfauthor={},
pdfsubject={},
pdfkeywords={},
pdfpagemode=UseThumbs,
]{hyperref}

\usepackage{orcidlink}

\newcommand{\bk}{{\bm{k}}}
\newcommand{\cm}{$\checkmark$}

\newcommand{\vj}{X}
\newcommand{\T}{$\mathcal{T}$\xspace}
\newcommand{\PT}{$\mathcal{PT}$\xspace}
\newcommand{\Pa}{$\mathcal{P}$\xspace}
\newcommand{\Real}[1]{\text{Re} \left[ #1 \right]}
\newcommand{\Imag}[1]{\text{Im} \left[ #1 \right]}

\begin{document}

\title{
Generalized Pitaevskii relation between rectifying and linear responses:\\
its application to reciprocal magnetization induction
}

\author{Hikaru Watanabe
        \orcidlink{0000-0001-7329-9638}~~} 
\email{hikaru-watanabe@g.ecc.u-tokyo.ac.jp}
\affiliation{Research Center for Advanced Science and Technology, University of Tokyo, {Meguro-ku}, Tokyo 153-8904, Japan}

\author{Akito Daido
        \orcidlink{0000-0003-4629-7818}~~} 
\address{Department of Physics, Graduate School of Science, Kyoto University, Kyoto 606-8502, Japan}

\begin{abstract}
        Nonlinear optics has regained attention in recent years, especially in the context of optospintronics and topological materials. Nonlinear responses involved in various degrees of freedom manifest their intricacy more pronounced than linear responses.
        However, for a certain class of nonlinear responses, a connection can be established with linear-response coefficients, enabling the exploration of diverse nonlinear-response functionality in terms of the linear-response counterpart.
        Our study quantum-mechanically elucidates the relation between such nonlinear and linear responses we call the Pitevskii relation and identifies the condition for the relation to hold.
        Following the obtained general formulation, we systematically identify the Pitaevskii relations such as the inverse magnetoelectric effect and inverse natural optical activity unique to systems manifesting the space-inversion-symmetry breaking.
        These results provide a systematic understanding of intricate nonlinear responses and may offer further implications to ultrafast spintronics.
\end{abstract}

\maketitle

\section{introduction}

Interaction between light and matter has been a subject of extensive research.
The interaction gives rise to a plethora of diverse physical phenomena stemming from the breaking of symmetry such as magneto-optical responses (\textit{e.g.}, Faraday effect).
In particular, the development in laser technology allows for high-intensity light sources and thereby makes it feasible to explore the optical responses of matter in more detail.
For instance, intense light induces nonlinear effects like the rectification phenomena driven by the oscillating stimuli.
The phenomena include rectified electric polarization (optical rectification)~\cite{Kleinman1962-rt,Bass1962-lf,Armstrong1962-sj}, magnetization (inverse Faraday and Cotton-Mouton effects)~\cite{Pitaevskii1961-rd,Armstrong1962-sj,Pershan1963-nd,van-der-Ziel1965-dm,Pershan1966-oa,Kalashnikova2007-yx,Hansteen2006-zg,Gridnev2008-ml,Kalashnikova2008-uj}, and mechanical rotation (Sadovskii effect)~\cite{Holbourn1936-mb,Beth1936-tb,Sokolov1991-lh} responses to the irradiating light.
The formula is explicitly given by
                \begin{equation*}
                M_i = \kappa_{ijk} (\omega) E_j (\omega) E_k (-\omega),
                \end{equation*}
for the rectified magnetization response $\bm{M}$ to the double electric field $\bm{E}^2$.

Further advancements in optical technology enable time-resolved spectroscopy of materials stimulated by the pulsive light, namely pump-probe spectroscopy.
Combined with the nonlinear light-matter interaction, the methodology has paved the way for ultrafast control of the phase of matter such as magnetism, garnering significant interest in the realm of ultrafast spintronics~\cite{Kirilyuk2010-gx,Nemec2018-wu,Stupakiewicz2021-gt}.
It is noteworthy that nonlinear light-matter coupling, being in sharp contrast to the thermalization-driven control~\cite{Beaurepaire1996-bu,Hohlfeld1997-nz,Kimel2004-xl,Duong2004-cj,Satoh2007-yr}, is expected to open a route to highly efficient manipulation of magnetic states with minimizing Joule dissipation~\cite{Kimel2005-vd,Stanciu2007-vr,Khorsand2012-vf,Kurkin2008-xh,Satoh2010-tt,Higuchi2011-qz}.

In the field of nonlinear magneto-optics, it has been confirmed that the non-absorption condition leads to nontrivial relationships bridging linear and nonlinear responses.
For instance, in the case of the Faraday effect, the response is determined by the off-diagonal elements of the optical dielectric susceptibility $\chi_{ij} (\omega)$, that is the optical Hall susceptibility.
As uncovered by Pitaevskii~\cite{Pitaevskii1961-rd,Landau2013electrodynamics}, the optical Hall susceptibility is correlated with the response function of the inverse Faraday effect $\kappa_{ijk}$ in the absence of the absorption as 
                \begin{equation*}
                \kappa_{ijk}  (\omega)= \frac{1}{2} \lim_{\bm{B} \to \bm{0}} \partial_{B_i} \, \chi_{jk} (\omega), 
                \end{equation*}
with the external magnetic field $\bm{B}$.
Here let us call the nontrivial relation between the linear and rectification responses \textit{Pitaevskii relation}.
The original derivation is carried out with the classical treatment of electrodynamics of matter and subsequently grounded in arguments based on effective free energies or perturbation analysis of the atom Hamiltonian~\cite{Pershan1963-nd,Pershan1966-oa}.
Furthermore, the effect of absorption has been elucidated in recent works including those for two-dimensional systems and first-principles calculations of bulk materials~\cite{Battiato2014-ex,Tokman2020-vl}.

Despite extensive research, the relation has hitherto not been formulated in a full-quantum manner to cover that between various degrees of freedom other than the electric and magnetic polarizations~\cite{Pershan1966-oa}.
Furthermore, the microscopic conditions for the Pitaevskii relation to hold remain elusive, though the absence of absorption is considered to be essential.
By considering the fact that the spintronic applications have been explored on the basis of theoretical and experimental findings delving into inverse magneto-optical effects~\cite{Kirilyuk2010-gx}, generalized Pitaevskii relations may allow us to explore crucial insights for future investigations of the optoelectronics.

In this study, we present a quantum derivation of Pitaevskii relations in a general formulation utilizing the Lehmann representation and auxiliary-field method~\cite{Watanabe2022-hk} without specific assumptions such as the independent-particle approximation.
Employing Kubo's response theory, we derive the conditions under which Pitaevskii relations hold among various degrees of freedom through calculations of linear and nonlinear responses.
We analytically and numerically demonstrate that Fermi surface effects may lead to the violation of Pitaevskii relations, even when the non-absorption condition is satisfied.
The obtained relations are systematically classified based on the system's space-time symmetry.
As an example, we extensively investigate the phenomena such as inverse magnetoelectric and inverse natural optical activity, which are unique to systems with broken space-inversion symmetry.

The organization is the following. 
In Sec.~\ref{SecSub_non-absorptive-relation}, we elucidate the non-absorption condition concerning the responses to the external stimuli up to the second order.
Performing full-quantum calculations, we derive the condition for the Pitaevskii relation to hold in Sec.~\ref{SecSub_Pitaevskii-relation_derivation}.
In Sec.~\ref{SecSub_symmetry_remarks}, after revisiting the Pitaevskii relation of the inverse Faraday effect, we generalize the relation to cover diverse correlations between the linear and rectification responses.
The systematic classification of the relations is presented in Sec.~\ref{SecSub_symmetry_T_PT_classification}, and a tabulation of them is in Sec.~\ref{SecSub_symmetry_tabulation}.
Our formulation is further demonstrated in Sec.~\ref{Sec_reciprocal_magnetization_induction} by taking the specific example, that is reciprocal magnetization induction.
Numerical results are in agreement with the analytical results.
Finally, we summarize the contents in Sec.~\ref{Sec_summary}.

\section{Quantum-mechanical formulation of Pitaevskii relation}
\label{Sec_Pitaevskii-relation_derivation}

We derive the relation between the rectification and linear responses by following Kubo's response theory.
We first discuss the non-absorption condition required to validate the Pitaevskii relation (Sec.~\ref{SecSub_non-absorptive-relation}).
After the perturbative calculations, we formulate the non-absorptive rectification and (generalized) Pockels responses and thereby derive the Pitaevskii relation (Sec.~\ref{SecSub_Pitaevskii-relation_derivation}).

\subsection{Non-absorption condition}
\label{SecSub_non-absorptive-relation}

The variation of total energy is given by the expectation value of the time-derivative of the total Hamiltonian $H(t)$ as
                \begin{equation}
                        \frac{d}{dt}\Braket{H(t) } = \text{Tr}\left(  H (t) \frac{d \rho (t)}{dt} + \rho (t) \frac{d H (t)}{dt}  \right),
                        \label{variation-energy}
                \end{equation}
where we introduced the density matrix $\rho (t)$ denoting the quantum state of the system.
Considering a closed system, one can see that the first term vanishes by the von Neumann equation $i d \rho (t) / dt = \left[ H, \rho (t) \right]$.
This indicates that the heat production is zero and the variation of energy is attributed to the work $W$ done by the external stimuli.
Then, we recast Eq.~\eqref{variation-energy} as
                \begin{equation}
                        \frac{d}{dt}\Braket{H(t) } = W \equiv \text{Tr}\left(  \rho (t) \frac{d H  (t)}{dt}  \right).
                \end{equation}
With the coupling between the stimuli and system written by $H_\text{ex} = \sum_i F_i (t) X_i+O(F^2)$, the work is given by 
                \begin{equation}
                W = \text{Tr}\left[  \rho (t) X_i (F(t)  \right]  \frac{dF_i(t)}{dt},\quad X_i(F(t))=\frac{\partial H_{\rm ex}}{\partial F_i(t)},
                \end{equation}
where the summation over the stimuli $i$ is implicit.
The response to the external stimuli is expanded with respect to $F$;
                \begin{align}
                \Braket{X } 
                        = \text{Tr}\left(  \rho (t) X_i(F(t))  \right) 
                        &= \Braket{X_i}_\text{eq} +  \int_{-\infty}^\infty dt' \chi_{ij} (t-t') F_j(t') \notag \\
                        &+ \int_{-\infty}^\infty dt' \int_{-\infty}^\infty dt'' \kappa_{ijk} (t-t',t-t'') F_j(t')F_k (t'') + O(F^3).
                \end{align}
$\Braket{\cdot }_\text{eq}$ denotes the averaging by the density matrix $\rho = \rho_\text{eq}$ under no external fields.
Note that each susceptibility tensor satisfies the causality as $\chi_{ij} (s) \propto \theta (s)$ and $\kappa_{ijk} (s,s') \propto \theta (s)\theta (s')$, where $\theta(s)$ is the Heaviside step function.

After subtracting the expectation value in equilibrium as $\Braket{X } \to \Braket{X } - \Braket{X }_\text{eq}$, the work is obtained in a perturbative manner.
It is given by
                \begin{equation}
                W^{(1)} = \int_{-\infty}^\infty dt' \partial_t F_i (t) \chi_{ij} (t-t')  F_j (t'),
                \end{equation}
up to the linear response.
We are interested in the work averaged over a sufficiently long period $T$ and therefore take the average of $W$
                \begin{equation}
                \overline{W}^{(1)} = \frac{1}{T}\int_{-T/2}^{T/2} dt W^{(1)} \to \int\frac{d\omega}{2\pi} i \omega F_i^\ast (\omega) \chi_{ij}^{(1)} (\omega) F_j (\omega).
                \label{energy-variation-linear}
                \end{equation}
On the rightmost side, we took the limit of $T\to \infty$.
We here define the non-absorption condition that the total work $\overline{W}$ done by the external stimuli $F(t)$ is zero,
                \begin{equation}
                \overline{W} = 0.
                \end{equation}
In Eq.~\eqref{energy-variation-linear}, the non-absorption condition is satisfied when the susceptibility tensor $\chi (\omega)$ is hermitian as $\chi_{ij}^\ast (\omega) = \chi_{ji} (\omega)$.
This can be proved by the fact that the stimuli $F_i  (t)$ is real and thereby $F_i^\ast (\omega) = F_i(-\omega)$.
For the case of dielectric susceptibility, the non-absorption condition is the vanishing imaginary part.

Similarly, we obtain the time-averaged work up to the second order.
The second-order correction is 
                \begin{align}
                \overline{W}^{(2)}  
                        &=\int\frac{d\omega d\omega_1 d\omega_2 }{(2\pi)^2} \delta (\omega - \omega_1 - \omega_2)  i \omega  F_i^\ast (\omega) \kappa_{ijk} (-\omega; \omega_1,\omega_2) F_j(\omega_1) F_k(\omega_2).
                        \label{second-order_energy-variation_not-symmetrized}
                \end{align}
We here defined the second-order susceptibility tensor in the frequency domain by
                \begin{equation}
                \kappa_{ijk} (-\omega_1-\omega_2; \omega_1,\omega_2) = \int dt dt'~ \kappa_{ijk} (s,s') e^{i\omega_1 s + i\omega_2 s'}.
                \end{equation}
$\overline{W}^{(2)}$ is rewritten as
                \begin{align}
                \overline{W}^{(2)}
                        &= \frac{1}{3}\int\frac{d\omega d\omega_1 d\omega_2 }{(2\pi)^2} \delta (\omega - \omega_1 - \omega_2) i F_i (\omega) F_j (\omega_1) F_k (\omega_2)  \left\{ \omega \kappa_{ijk} (-\omega;\omega_1,\omega_2) - \omega_1 \kappa_{jik} (\omega_1;-\omega,\omega_2)- \omega_2 \kappa_{kij} (\omega_2;-\omega,\omega_1) \right\},\\ 
                        &= -\frac{1}{3}\int\frac{d\omega d\omega_1 d\omega_2 }{(2\pi)^2}  i F_i (\omega) F_j (\omega_1) F_k (\omega_2) \tilde{\kappa}_{ijk} (-\omega,\omega_1,\omega_2).
                \end{align}
In the final line, we defined the tensor
                \begin{equation}
                        \tilde{\kappa}_{ijk} (\omega_i,\omega_j,\omega_k) = \delta (\omega_i + \omega_j + \omega_k) \left\{ \omega_i \kappa_{ijk} (\omega_i;\omega_j,\omega_k) + \omega_j \kappa_{jik} (\omega_j;\omega_i,\omega_k)+ \omega_k \kappa_{kij} (\omega_k;\omega_i,\omega_j) \right\}.\label{eq:kappa_prime}
                \end{equation}
Owing to the intrinsic permutation symmetry of the second-order susceptibility tensor, that is $\kappa_{ijk} (\omega_i;\omega_j,\omega_k) = \kappa_{ikj} (\omega_i;\omega_k,\omega_j)$, the tensor $\tilde{\kappa}_{ijk} (\omega_i,\omega_j,\omega_k)$ is totally symmetric under any permutation between $(i,\omega_i)$, $(j,\omega_j)$, and $(k,\omega_k)$.
As a result, the second-order averaged work $\overline{W}^{(2)}$ vanishes when the totally symmetric tensor $\kappa_{ijk} (\omega_i,\omega_j,\omega_k)$ is zero.

The vanishing second-order correction is related to the full permutation symmetry~\cite{Boyd2020-rc}, that is
                \begin{equation}
                        \kappa_{ijk} (\omega_i;\omega_j,\omega_k) = \kappa_{jik} (\omega_j;\omega_i,\omega_k) = \kappa_{kij} (\omega_k;\omega_i,\omega_j).
                \end{equation}
When the full-permutation symmetry holds, one can straightforwardly find that the totally symmetric part is zero; $\tilde{\kappa}_{ijk} (\omega_i,\omega_j,\omega_k) =0$.
Note, on the other hand, that the full permutation symmetry may not always be satisfied when the totally-symmetric tensor $\tilde{\kappa}_{ijk} (\omega_i,\omega_j,\omega_k)$ is zero. 
Thus, the full permutation symmetry is a sufficient condition for the non-absorption, though is not necessary.

Let us elaborate on the non-absorption condition and the full permutation symmetry in the case of the rectification response, which is of primary interest.
They are given by,
                \begin{equation}
                \tilde{\kappa}_{ijk} (0,-\omega,\omega) = 0,
                \label{rectification_non-absorption}
                \end{equation}
and
                \begin{equation}
                \kappa_{ijk} (0;-\omega,\omega) = \kappa_{jik} (-\omega;0,\omega) = \kappa_{kij} (\omega;0,-\omega).
                \label{rectification_non-absorption_full-permutation}
                \end{equation}
The full permutation symmetry indicates that the rectification response is related to the Pockels effect denoted by $\kappa_{jik} (-\omega;0,\omega)$ and $\kappa_{kij} (\omega;0,-\omega)$, that is DC-field ($F_i$) correction to the linear susceptibility concerning $F_j (-\omega)$ and $F_k (\omega)$.
Note that the Pockels effect is for the DC-electric-field correction to the electric permittivity, which is reproduced by taking $(F_i, F_j, F_k) = (E_p, E_q, E_r)$, where the indices $p$, $q$, and $r$ are for the real-space coordinates.
In the following, we call the second-order responses of the form $\kappa_{ijk}(\pm\omega;0,\mp\omega)$ Pockels effects regardless of whether the external stimuli is the electric field.

In closing this section, we note that the stimulus $F_i$ in the above discussion is assumed to be invariant under the gauge transformation.
For instance, the electric current $\bm{J}$ is conjugated to the vector potential $\bm{A}$, and thus in the case of the electric current response to the quadratic electric field, the fields are taken as $(F_i, F_j, F_k) = (A_p, E_q, E_r)$ with the electric field $\bm{E}$ in Eq.~\eqref{second-order_energy-variation_not-symmetrized}.
In this case, it is more physically transparent to replace the vector potential with the electric field as $\bm{A} (\omega) = \bm{E}(\omega) /(i\omega)$, to make the gauge invariance of $\overline{W}^{(2)}$ manifest.
Then, we rewrite the second-order contribution to the averaged work $\overline{W}$ as
                \begin{align}
                \overline{W}_{JPP}^{(2)}  
                        &=\int\frac{d\omega d\omega_1 d\omega_2 }{(2\pi)^2} \delta (\omega - \omega_1 - \omega_2)   \kappa'_{ijk} (-\omega; \omega_1,\omega_2) E_i^\ast (\omega) E_j(\omega_1) E_k(\omega_2).
                        \label{second-order_energy-variation_PPtoJ}
                \end{align}
Here $P$ in the subscript of $\overline{W}_{JPP}^{(2)}$ denotes the electric polarization conjugate to the electric field.
Accordingly, the totally-symmetric tensor defined from $\kappa'$ is given by
                \begin{equation}
                        \tilde{\kappa}'_{ijk} (\omega_i,\omega_j,\omega_k) = \delta (\omega_i + \omega_j + \omega_k) \left\{ \kappa'_{ijk} (\omega_i;\omega_j,\omega_k) + \kappa'_{jik} (\omega_j;\omega_i,\omega_k)+ \kappa'_{kij} (\omega_k;\omega_i,\omega_j) \right\},
                        \label{totally-symmetric-tensor_current-representation}
                \end{equation}
with which finite work is obtained through the second-order current response.
The obtained non-absorption condition $\tilde{\kappa}'_{ijk}=0$ is consistent with Refs.~\cite{Shi2023-fr} and \cite{Tsirkin2022-zf} where the DC current response to the AC and DC electric fields are discussed, respectively.
In this way, when the output is conjugate to the gauge-covariant field, one can discuss the non-absorption condition with the totally-symmetric tensor such as $\tilde{\kappa}'_{ijk} (\omega_i,\omega_j,\omega_k)$ of Eq.~\eqref{totally-symmetric-tensor_current-representation} instead of $\tilde{\kappa}_{ijk}(\omega_i,\omega_j,\omega_k)$ of Eq.~\eqref{eq:kappa_prime}.

\subsection{Derivation of Pitaevskii relation}
\label{SecSub_Pitaevskii-relation_derivation}

We formulate the rectification and the Pockels effects.
Following the established perturbative calculation~\cite{Watanabe2022-hk,de-Juan2020-ev,Watanabe2021-bt,Ahn2020-ec,Passos2018-cc,Michishita2021-rb,Holder2020-tv,von-Baltz1981-yf,Sipe2000-sb,Morimoto2016-ne,Ventura2017-db,Oiwa2022-he}, we obtain the nonlinear susceptibility in the Lehmann representation as
        \begin{equation}
                    X^i_{(2)} (\omega)
                     = \int \frac{d\omega_1 d\omega_2 }{(2\pi)^2} 2\pi \delta (\omega -\omega_1 - \omega_2) \kappa_{ijk} (-\omega;\omega_1,\omega_2) F_{j}(\omega_1) F_{k} (\omega_2), \label{RDM_full_2nd_optical_conductivity}
        \end{equation}
where 
        \begin{align}
        \kappa_{ijk}(-\omega;\omega_1,\omega_2)
            &=\lim_{\bm{F}\to \bm{0}} \Biggl[ \sum_a \frac{1}{2}  \vj^{ijk}_{aa} f_a \label{RDM_2;0} \\ 
             &+\sum_{a,b} \frac{1}{2} \frac{\vj^{ij}_{ab} \vj^k_{ba} f_{ab} }{ \omega_2 +i\eta -\epsilon_{ba}} + \frac{1}{2} \frac{\vj^{ik}_{ab} \vj^j_{ba} f_{ab} }{ \omega_1 +i\eta -\epsilon_{ba}}\label{RDM_1;1} \\
             &+\sum_{a,b}\frac{1}{2}  \frac{\vj^{i}_{ab} \vj^{jk}_{ba} f_{ab} }{ \omega + 2i\eta -\epsilon_{ba}}\label{RDM_0;2_2photon}\\
             &+\sum_{a,b,c} \frac{1}{2}\frac{\vj^{i}_{ab}}{ \omega + 2i\eta -\epsilon_{ba}} \left(   \frac{\vj^j_{bc}\vj^{k}_{ca}f_{ac} }{\omega_2 + i \eta -\epsilon_{ca} }    -    \frac{\vj^j_{ca}\vj^{k}_{bc}f_{cb} }{\omega_2 + i \eta -\epsilon_{bc} }  \right)\label{RDM_0;2_1photon_1}\\
             &+\sum_{a,b,c}\frac{1}{2}\frac{\vj^{i}_{ab}}{ \omega + 2i\eta -\epsilon_{ba}} \left(  \frac{\vj^k_{bc}\vj^{j}_{ca}f_{ac} }{\omega_1 + i \eta -\epsilon_{ca} }    -    \frac{\vj^k_{ca}\vj^{j}_{bc}f_{cb} }{\omega_1 + i \eta -\epsilon_{bc} } \right)  \Biggr]. \label{RDM_0;2_1photon_2}
       \end{align}
Here, $\eta=+0$ represents the adiabaticity parameter.
We defined the operators 
                \begin{align}
                \vj^{i} &= \left. \frac{\partial H (\bm{F})}{\partial F_i} \right.,\\
                \vj^{ij} &=\left. \frac{\partial^2 H (\bm{F})}{\partial F_i\partial F_j}\right.,\\
                \vj^{ijk} &= \left.\frac{\partial^3 H(\bm{F})}{\partial F_i\partial F_j\partial F_k}\right.,
                \end{align}
where the Hamiltonian $H(\bm{F})$ includes the coupling to the stimuli.
In the case of the vector potential ($\bm{F} = \bm{A}$), for instance, $\vj^i$ and $\vj^{ij}$ denote the paramagnetic and diamagnetic current operators in the limit of $\bm{F} \to \bm{0}$, respectively.
We also introduced the energy eigenvalue $\epsilon_{a}$ for the many-body Hamiltonian including $H_\text{ex}$ and the Boltzmann factor $f_a = e^{-\epsilon_a /T} /\left( \sum_b e^{-\epsilon_b /T} \right)$, and accordingly defined $\epsilon_{ab} = \epsilon_{a}-\epsilon_{b}$ and $f_{ab} = f_a - f_b$.
While we here consider general interacting systems, we can show for non-interacting electron systems that the equations of the same form as the following ones hold by replacing $H(\bm{F})$ in the definition of $X^i$, $X^{ij}$ and $X^{ijk}$ with the single-particle Hamiltonian and accordingly reinterpreting energy eigenstates and eigenvalues, as well as replacing $f_a$ with the Fermi distribution function.

First, we consider the rectification response $\kappa_{ijk} (0;-\omega,\omega)$.
Following the parallel discussions in Ref.~\cite{Watanabe2022-hk}, we arrive at the expression including no resonant contribution;
        \begin{align}
        \kappa_{ijk}^\text{na} (0;-\omega,\omega)
                =\frac{1}{2} \lim_{\bm{F}\to \bm{0}} \left\{\partial_{F_i} \left[ \sum_{a}  \vj^{jk}_{aa}  f_a  - \sum_{a,b} \vj^j_{ba} \vj_{ab}^k f_{ab} \frac{1}{\omega + \epsilon_{ba}} \right] - \sum_a \vj^{jk}_{aa} \partial_{F_i} f_a - \sum_{a,b} \vj^j_{ab} \vj_{ba}^k  \frac{1}{\omega + \epsilon_{ab}}  \partial_{F_i}  f_{ab} \right\}.
                \label{off-resonant_rectification-response}
        \end{align}
Similarly, the off-resonant Pockels response is given by
        \begin{equation}
        \kappa_{ijk}^\text{na} (\omega;-\omega,0)
                =\frac{1}{2} \lim_{\bm{F}\to \bm{0}} \left\{\partial_{F_k} \left[ \sum_{a}  \vj^{ij}_{aa}  f_a  - \sum_{a,b} \vj^j_{ba} \vj_{ab}^i f_{ab} \frac{1}{\omega + \epsilon_{ba}} \right] -\sum_a \vj^{ij}_{aa} \partial_{F_k} f_a - \sum_{a,b} \vj^j_{ab} \vj_{ba}^i  \frac{1}{\omega + \epsilon_{ab}}  \partial_{F_k}  f_{ab} \right\}.
                \label{off-resonant_Pockels}
        \end{equation}
Here we divided the total second-order susceptibility into terms with and without resonant contributions by $\kappa_{ijk}(\omega_i;\omega_j,\omega_k)=\kappa_{ijk}^{\rm a}(\omega_i;\omega_j,\omega_k)+\kappa_{ijk}^{\rm na}(\omega_i;\omega_j,\omega_k)$.
The resonant contribution $\kappa_{ijk}^{\rm a}$ is defined to include delta functions that appear from factors like $(\omega+i\eta-\epsilon)^{-1}$.
The derivations are given in Appendix~\ref{SecApp_derivation_off-resonant_rectification}.
It turns out that $\kappa^\text{na}$ is responsible for the Pitaevskii relation and thus we focus on this component.

The obtained response functions satisfy
                \begin{equation}
                        \kappa_{ijk}^\text{na} (0;-\omega,\omega) = \kappa_{kji}^\text{na} (\omega;-\omega,0).
                \end{equation}
One can straightforwardly derive other relations such as $\kappa_{ijk}^\text{na} (0;-\omega,\omega) = \kappa_{jik}^\text{na} (-\omega;0,\omega)$ by using the intrinsic permutation symmetry.
These relations indicate that the full-permutation symmetry of Eq.~\eqref{rectification_non-absorption_full-permutation} is satisfied for $\kappa^{\rm na}_{ijk}$.
Thus, we conclude that the off-resonant rectification- and Pockels-response functions satisfy the non-absorption condition.
The response functions \eqref{off-resonant_rectification-response} and~\eqref{off-resonant_Pockels} consist of two contributions; the first term enclosed by auxiliary-field derivative and the second term including the distribution-modulation effect ($\partial_{\bm{F}}f_a$).

Linear-response functions are similarly obtained as 
                \begin{equation}
                \chi_{ij} (\omega) = \sum_a \vj_{aa}^{ij} f_a + \sum_{a,b} \frac{\vj_{ab}^i\vj_{ba}^j}{\omega + i\eta +\varepsilon_{ab}}f_{ab}.
                \label{linear_susceptibility_function}
                \end{equation}  
The non-absorptive part is given by the hermitian component 
                \begin{equation}
                \chi_{ij}^\text{na} (\omega) = \frac{1}{2} \left\{ \chi_{ij} (\omega) + \chi_{ji}^\ast (\omega) \right\} = \sum_a \vj_{aa}^{ij} f_a +\sum_{a,b}\mathrm{P} \frac{\vj_{ab}^i\vj_{ba}^j }{\omega - \epsilon_{ba}}f_{ab},
                \label{linear-response_non-absorption}
                \end{equation}
where $\mathrm{P}$ denotes the Cauchy principal value.
This means that the non-absorption condition for the linear response means the absence of resonant contributions.
Finally, we can relate the rectification response with the linear response as
                \begin{equation}
                        \kappa_{ijk}^\text{na} (0;-\omega,\omega) = \frac{1}{2} \lim_{\bm{F}\to \bm{0}} \left\{\partial_{F_i} \chi_{jk}^\text{na} (-\omega,\bm{F}) +\kappa_{ijk}^\text{dm}  \right\}.
                        \label{pitaevskii-relation_with_fermi-surface_factor}
                \end{equation}
The auxiliary-field ($\bm{F}$) dependence of the linear-response function is explicitly shown as $\chi_{ij} (\omega,\bm{F})$ and 
                \begin{equation}
                        \kappa_{ijk}^\text{dm} \equiv - \sum_a \vj^{jk}_{aa} \partial_{F_i} f_a - \sum_{a,b} \vj^j_{ab} \vj_{ba}^k  \frac{1}{\omega + \epsilon_{ab}}  \partial_{F_i}  f_{ab},
                        \label{rectification_fermi-surface_factor}
                \end{equation}
is the contributions including the modulation of the distribution function.
Thus, we have proved that an equation similar to the Pitaevskii relation generally holds between off-resonant contributions of linear and nonlinear susceptibilities.
The Pitaevskii relation holds as
                \begin{equation}
                        \kappa_{ijk} (0;-\omega,\omega) = \frac{1}{2} \lim_{\bm{F}\to \bm{0}} \partial_{F_i} \chi_{jk} (-\omega,\bm{F}),
                        \label{eq:Pitaevskii-relation}
                \end{equation}
when there is neither $\kappa^{\rm dm}_{ijk}$ nor the resonant contributions of the linear and nonlinear susceptibilities, which are given by $\kappa^{\rm a}_{ijk}(0;-\omega,\omega)$ and the anti-hermitian part $\chi_{ij}^\text{a} (\omega) = (\chi_{ij}(\omega)-\chi^*_{ji}(\omega))/2$, respectively.

In summary, the full-quantum derivation of the rectification and Pockels effects clarified the condition for the Pitaevskii relation to hold beyond arguments based on the steady-state free energy~\cite{Pershan1963-nd} and on the atom Hamiltonian~\cite{Pershan1966-oa}.
Our formulation is based on the Lehmann representation of the response functions without considering specific approximations such as independent-particle approximation.
Pitaevskii relation~\eqref{eq:Pitaevskii-relation} holds if and only if the frequency $\omega$ is in the off-resonant regime and the distribution-modulation factor is negligible ($\partial_{\bm{F}} f_a = 0$); \textit{i.e.}, the rectification, Pockels, and linear responses are related with each other when interband-like or intraband-like excitation is absent.

It is noteworthy that the non-absorption condition does not always ensure the Pitaevskii relation.
For example, let us take the band-electron system where electrons partially occupy a single band well isolated from other bands.
If the frequency of the external field is sufficiently larger than the bandwidth but does not give rise to interband transition, the rectification-response and linear-response functions are given by the non-absorptive contributions of Eqs.~\eqref{rectification_non-absorption},~\eqref{linear-response_non-absorption}, whereas Pitaevskii relation is violated due to the Fermi-surface effect of Eq.~\eqref{rectification_fermi-surface_factor}.
Note that the summation over the eigenstates may result in the vanishing distribution-modulation effect in some situations~\cite{Belinicher1986-tk,Onishi2022-do}.
In the following subsections, we will mainly work on cases satisfying the non-absorption conditions ($\chi_{ij}^\text{a} = 0 $, $\kappa_{ijk}^\text{a} = 0 $) as well as $\kappa_{ijk}^\text{dm} =0$ to corroborate Pitaevskii relations.
Then, the superscript `na' will be suppressed unless explicitly mentioned.

\section{Generalized Pitaevskii relation and symmetry}
\label{Sec_symmetry}

\subsection{General remarks}
\label{SecSub_symmetry_remarks}

Our formulation covers diverse Pitaevskii relations including known results for the inverse Faraday, inverse Cotton-Mouton, and optical rectification effects.
One can take various fields for each auxiliary field such as electric field $\bm{E}$, spin and orbital Zeeman field $\bm{B}_\text{sp/orb}$, stress $\sigma_{ij}$, and so on.
Furthermore, the auxiliary field may be taken as what is related to the spontaneous symmetry breaking such as the spatial gradient of the phase of the condensate of Cooper pairs~\cite{Mironov2021-gq,Watanabe2022-hk,Watanabe2022-gh}, which is equal to the vector potential $\bm{A}$ in the London gauge.

We here investigate cases where the Pitaevskii relation holds as 
                \begin{equation}
                        \kappa_{ijk} (0;-\omega,\omega) = \frac{1}{2} \lim_{\bm{F}\to \bm{0}} \partial_{F_i} \chi_{jk} (-\omega,\bm{F}).
                        \label{generalized_Pitaevskii-relation}
                \end{equation}
The distribution-modulation contribution is assumed to be zero.
Pitaevskii relations are classified by the preserved symmetry of the unperturbed Hamiltonian.
Let us take a known example, that is the magnetization response to the double electric field.
The induced magnetization is 
                \begin{equation}
                M_i = \kappa_ {ijk}^{BEE} (\omega)E_j  (-\omega) E_k (\omega) = \kappa_{ijk}^{BEE} (\omega) E_j^\ast  (\omega) E_k (\omega),
                \label{EE_to_magnetization}
                \end{equation}
where we explicitly show the auxiliary fields $(F_i,F_j,F_k) = (\bm{B},\bm{E},\bm{E})$ related to the response in the superscripts of the susceptibility $\kappa_{ijk} (\omega)\equiv\kappa_{ijk}(0;-\omega,\omega)$.
According to the Pitaevskii relation, $\kappa_{ijk}^{BEE}$ is related to the electric permittivity defined by the formula $P_j (\omega) = \chi_{jk}^{EE} (\omega) E_k (\omega)$.
The Pitaevskii relation is explicitly written as
                \begin{equation}
                        \kappa_{ijk}^{BEE} (\omega) = \frac{1}{2} \lim_{\bm{B}\to 0}\partial_{B_i} \chi_{jk}^{EE} (-\omega,\bm{B}).
                        \label{EE_to_magnetization_pitaevskii}
                \end{equation}

The anti-unitary symmetry such as the time-reversal (\T{}) symmetry is convenient to decompose the relation.
When the \T{}-symmetry is intact in the unperturbed state, the Onsager reciprocity relation leads to
                \begin{equation}
                        \chi_{jk}^{EE} (\omega,\bm{B}) = \chi_{kj}^{EE} (\omega,-\bm{B}),
                        \label{Onsager_electric-permittivity}
                \end{equation}
by which only the antisymmetric component ($\chi_{jk}^{EE} = -\chi_{kj}^{EE}$) contributes to the Pitaevskii relation of Eq.~\eqref{EE_to_magnetization_pitaevskii}.
It follows that the DC magnetization is induced by the cross-product of the double electric fields;
                \begin{equation}
                        M_i = \frac{1}{2} \kappa_{ijk}^{BEE} (\omega)~ \epsilon_{jkl} \left( \bm{E}^\ast  (\omega)\times \bm{E} (\omega) \right)_l,
                \end{equation}  
in \T{}-symmetric systems.
Since $\bm{E}^\ast  (\omega)\times \bm{E} (\omega) \neq \bm{0}$ when the light has the circular-polarized component~\cite{Wolf2007Book_coherence}, the obtained formula represents the DC magnetization response to the circularly-polarized light, so-called the inverse Faraday effect~\cite{Pitaevskii1961-rd,van-der-Ziel1965-dm}.
As a result, the Pitaevskii relation of Eq.~\eqref{EE_to_magnetization_pitaevskii} points to the correlation between the inverse Faraday effect and optical Hall conductivity.
When the \T{} symmetry is not kept in the unperturbed Hamiltonian, the Onsager reciprocity relation [Eq.~\eqref{Onsager_electric-permittivity}] does not hold.
Then, DC magnetization can respond to the non-circular part of the double electric fields (unpolarized and linearly-polarized lights) satisfying $E_j^\ast  (\omega) E_k (\omega) = E_k^\ast  (\omega) E_j (\omega)$, that is the inverse Cotton-Mouton effect~\cite{Kalashnikova2007-yx,Hansteen2006-zg}~\footnote{
    The Cotton-Mouton effect is magnetic birefringence proportional to the square of magnetization $\bm{M}$.
    We note that this magnetic birefringence is attributed not only to the magnetic correction to the symmetric part of the electric permittivity ($\Delta \chi_{ij}^{EE} = +\Delta \chi_{ji}^{EE} \propto \bm{M}^2$) but also that to the antisymmetric part ($\Delta \chi_{ij}^{EE} = - \Delta \chi_{ji}^{EE} \propto \bm{M}$) in the Voigt optical arrangement.
    We here consider the Cotton-Mouton in a narrow sense, that is the magnetic birefringence arising from corrections to the symmetric part. 
}.

Note that the permutation symmetry between the indices of double electric fields $(j,k)$ is in close relation to the property as the complex number; the antisymmetric part ($\bm{E}^\ast  (\omega)\times \bm{E} (\omega) $) is pure imaginary while the symmetric part ($E_j^\ast  (\omega) E_k (\omega) + (j\leftrightarrow k)$) is real.
Accordingly, the rectification-response function $\kappa_{ijk}^{BEE}$ is divided into the imaginary and real parts for the inverse Faraday and Cotton-Mouton effects, respectively.
The decomposition is explicitly given as
                \begin{align}
                        M_i 
                                &= \frac{1}{2} \Real{\kappa_{ijk}^{BEE} (\omega)}  \left\{ E_j^\ast  (\omega) E_k (\omega) + \left( j \leftrightarrow k \right)  \right\} + \frac{i}{2} \Imag{\kappa_{ijk}^{BEE}(\omega)} \left\{ E_j^\ast  (\omega) E_k (\omega) - \left( j \leftrightarrow k \right)  \right\}.
                \end{align}
If the non-absorption condition is satisfied, the real-imaginary decomposition of $\kappa_{ijk}^{BEE}$ is consistent with the symmetry of the non-absorptive linear response whose response function is hermitian ($\chi_{ij}^\ast(\omega) = \chi_{ji}(\omega)$), and thereby the permutation symmetry of indices determines whether the response function is real or purely imaginary as
                \begin{equation}
                \Real{\chi_{ij} (\omega)} = \Real{ \chi_{ji} (\omega)}, ~ \Imag{ \chi_{ij} (\omega)}  = - \Imag{\chi_{ji} (\omega)}.
                \end{equation}
Note that the symmetry of the Pitaevskii relation does not change when the electric fields are replaced by the magnetic fields in Eq.~\eqref{EE_to_magnetization}~\cite{Takayoshi2014-wi} or when the induced magnetization is replaced with another magnetic multipolar degree of freedom having the ferromagnetic symmetry such as magnetic octu-polarization in Mn$_3$Sn~\cite{Suzuki2017-ps}.

The classification is straightforwardly generalized.
Let us take the \T{}-symmetric system and define the \T{} parity of the physical field $F_i$ by $\tau_\theta^{F_i}$.
The Onsager reciprocity relation reads as
                \begin{equation}
                \chi_{jk} (\omega,F_i) = \tau_\theta^{F_j}\tau_\theta^{F_k}\chi_{kj} (\omega, \tau_\theta^{F_i} F_i),
                \label{Onsager-relation_general-case_T-parity}
                \end{equation}
due to which only the antisymmetric double fields ($\epsilon_{jkl} F_j^\ast F_k$) is relevant to Pitaevskii relation if the total parity is odd as $\tau_\theta^\text{tot} \equiv \tau_\theta^{F_i}\tau_\theta^{F_j}\tau_\theta^{F_k} = - 1 $.
Conversely, only the symmetric components ($ F_j^\ast F_k + (j \leftrightarrow k)$) are relevant if the total parity is even as $\tau_\theta^\text{tot} =+ 1 $.
Once the \T{} symmetry is violated in the unperturbed state, the symmetric and anti-symmetric parts also make contributions to the Pitaevskii relation for the cases of $\tau_\theta^\text{tot} = - 1 $ and $\tau_\theta^\text{tot} = + 1 $, respectively.
One can reproduce the symmetry of the DC magnetization response to the double electric fields from the above-mentioned classification by taking $\tau_\theta^\text{tot} = \tau_\theta^{\bm{H}} (\tau_\theta^{\bm{E}})^2 = -1$. 
We summarize the symmetry of the Pitaevskii relation in Table~\ref{Table_classification_pitaevkii_T_symmetry}~\footnote{
        One can find the correspondence between the permutation symmetry of indices and \T{} parity (Table~\ref{Table_classification_pitaevkii_T_symmetry}) even when including Eq.~\eqref{rectification_fermi-surface_factor} as in Eq.~\eqref{pitaevskii-relation_with_fermi-surface_factor}.
        Note that, if one adopts Eq.~\eqref{pitaevskii-relation_with_fermi-surface_factor}, the Pitaevskii relation does not hold.
}.

                \begin{table}[htbp]
                \caption{
                        Classification of Pitevskii relation based on the total time-reversal parity.
                        The non-absorptive rectification-response function $\kappa_{ijk}^\text{na}$ is classified by the permutation symmetry for $(j,k)$ and the total time-reversal parity $\tau_\theta^\text{tot}$.
                        The symmetric and antisymmetric parts are real and pure imaginary, respectively.
                        }
                \label{Table_classification_pitaevkii_T_symmetry}
                \centering
                \renewcommand{\arraystretch}{2}
                \begin{tabular}{lcc}
                 \toprule
                 & Symmetric ($\kappa_{ijk}^\text{na} = \kappa_{ikj}^\text{na}$)& Anti-symmetric ($\kappa_{ijk}^\text{na} =- \kappa_{ikj}^\text{na}$)\\
                 \midrule
                 $\tau_\theta^\text{tot} =+1$ & \T{}-even & \T{}-odd\\
                 $\tau_\theta^\text{tot} =-1$ & \T{}-odd & \T{}-even\\
                 \bottomrule
                \end{tabular}
                \end{table}

\subsection{Odd-parity response and \T{}-\PT{} classification}
\label{SecSub_symmetry_T_PT_classification}

Every system satisfies the \T{} symmetry if there is no external field or spontaneous symmetry breaking.
Among the responses classified in Table~\ref{Table_classification_pitaevkii_T_symmetry}, \T{}-even responses may exist in general, while the \T{}-odd contributions are admixed with them only when the \T{} symmetry is lost in the unperturbed state.
On the other hand, the \T{}-odd response can occur without admixed with the \T{}-even part if another symmetry forbids the latter.
For instance, the combined symmetry for the \T{} and the space-inversion (\Pa{}) operations, namely \PT{} symmetry, is convenient to classify the physical phenomena induced by the \Pa{}-breaking effect~\cite{Watanabe2024-ju}.
Similarly to Eq.~\eqref{Onsager-relation_general-case_T-parity}, the \PT{} symmetry leads to a kind of the Onsager reciprocity relation of the linear-response function as
                \begin{equation}
                        \chi_{jk} (\omega,F_i) = \tau_{\theta I}^{F_j}\tau_{\theta I}^{F_k}\chi_{kj} (\omega, \tau_{\theta I}^{F_i} F_i),
                \label{Onsager-relation_general-case_PT-parity}
                \end{equation} 
with $\tau_{\theta I}^F$ denoting the \PT{} parity.
Since $\tau_{\theta I}^F = \tau_\theta^F \cdot \tau_I^F $ ($\tau_I^F$ is the \Pa{} parity), the \PT{}-symmetry constraint on the linear response is contrasting to that demanded by the \T{} symmetry if one consider the odd-parity response in which $\tau_\theta^\text{tot} = - \tau_{\theta I}^\text{tot}$.
For instance, the symmetric part $\kappa_{ijk} = \kappa_{ikj}$ is allowed in \T{}-symmetric systems but is absent in the \PT{}-symmetric systems if $\tau_\theta^\text{tot} = +1$ and $\tau_{\theta I }^\text{tot} = -1$, whereas it is forbidden by the \T{} symmetry but can be finite in the \PT{}-symmetric if $\tau_\theta^\text{tot} = -1$ and $\tau_{\theta I }^\text{tot} = +1$.
Following the parallel discussions, we obtain the \T{}-\PT{} classification of the anti-symmetric part ($\kappa_{ijk} = - \kappa_{ikj}$).
One may obtain a similar classification by making use of the combined operation of the \T{} and another unitary operation such as $\theta 2$ comprised of the two-fold rotation operation, which may work for the classification of \Pa{}-even responses as well.

\begin{figure}[htbp]
        \centering
        \includegraphics[width=0.750\linewidth,clip]{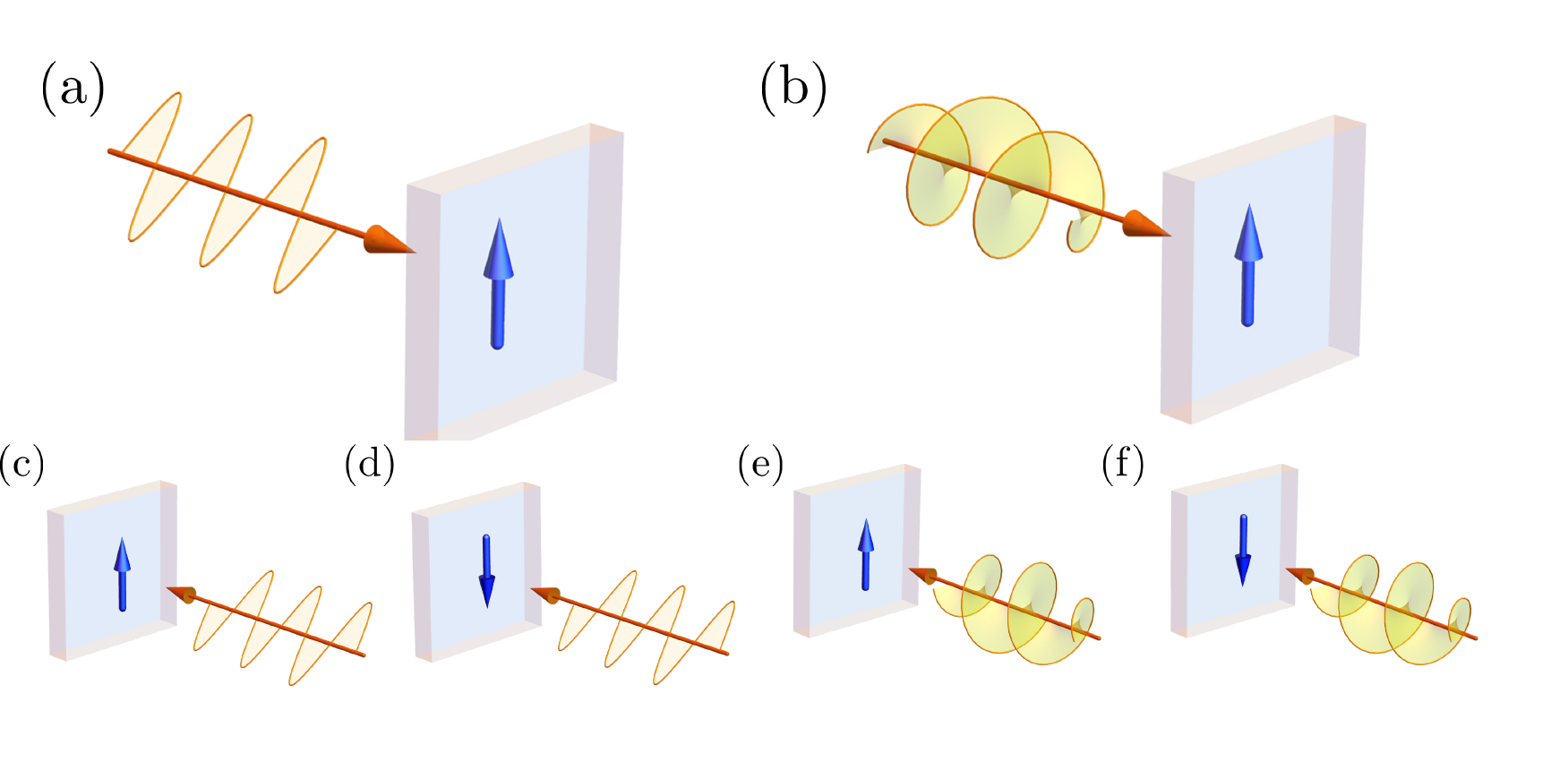}
        \caption{
                DC magnetization responses to (a) unpolarized or linearly-polarized light and (b) circularly-polarized light.
                Blue arrows are the induced magnetization and orange arrows are the propagating electromagnetic fields.
                The induced magnetization is not flipped and flipped under inversion of incident light for the \Pa{}-even and \Pa{}-odd rectification responses, respectively.
                In terms of reciprocity, panels are for (c) inverse Cotton-Mouton effect, (d) inverse magnetoelectric effect, (e) inverse Faraday effect, and (f) inverse natural optical activity.
        }
        \label{Fig_photo_induced_magnetization}
        \end{figure}

Let us consider an example where the \T{} and \PT{} symmetries play contrasting roles.
To this end, we consider an odd-parity rectification response written by
                \begin{align}
                M_i  
                    &= \kappa_{ijk}^{BEB} (\omega) E_j^\ast  (\omega) B_k (\omega) + \kappa_{ijk}^{BBE} (\omega) B_j^\ast  (\omega) E_k (\omega),\\
                \label{EH_to_magnetization}
                    &\equiv \left( \hat{\kappa}_{i}^{BEB} (\omega) \right)_{jk}  E_j^\ast  (\omega) B_k (\omega) + \left( \hat{\kappa}_{i}^{BBE} (\omega) \right)_{jk} B_j^\ast  (\omega) E_k (\omega),
                \end{align}
that is the DC magnetization response to the bilinear product of electric and magnetic fields.
The induced magnetization is flipped under inversion of the incident light, different from the response of Eq.~\eqref{EE_to_magnetization}.
Thus, the response formula denotes the \textit{reciprocal magnetization induction} (RMI).
We compare the response with the known $\bm{E}$-induced DC magnetization of Eq.~\eqref{EE_to_magnetization} in Fig.~\ref{Fig_photo_induced_magnetization} where
we make use of Faraday's law for the monochromatic field ($\bm{B} = \bk \times \bm{E}/\omega$). 
RMI may be overwhelmed by the inverse Faraday and Cotton-Mouton effects since the photo-magnetic field is typically smaller than the photo-electric field.
It is evident from the fact that its experimental observation remains elusive~\cite{Atzori2021-la}.
The observation, however, may be feasible by careful estimation of the magnetization induced by the lights propagating in the forward and backward directions.

To discuss the Pitaevskii relation, we rewrite the formula for RMI by
                \begin{align}
                M_i
                &= C_j^\ast (\omega) \Theta_{ijk} (\omega) C_k (\omega)\\
                &\equiv \bm{C}^\dagger (\omega) 
                \begin{pmatrix}
                        O &  \hat{\kappa}_{i}^{BEB} (\omega)\\       
                        \hat{\kappa}_{i}^{BBE} (\omega) & O\\       
                \end{pmatrix}
                \bm{C}(\omega),
                \end{align}
with the $3\times 3$ zero matrix $O$ and $\bm{C} = \left( \bm{E} (\omega),\bm{B} (\omega) \right)^T$.
The indices $i=1,2,3$ and $j,k = 1,2,\cdots,6$ for the response function $\Theta_{ijk}$.
Then, the Pitaevskii relation for RMI is 
                \begin{align}
                        \Theta_{ijk} (\omega) 
                                &= \frac{1}{2} \lim_{\bm{B}\to 0}\partial_{B_i} \chi_{jk}^{CC} (-\omega,\bm{B}), 
                        \label{EH_to_magnetization_pitaevskii}
                \end{align}
or equivalently
                \begin{equation}
                    \kappa_{ijk}^{BEB}(\omega)=\frac{1}{2}\lim_{\bm{B\to0}}\partial_{B_i}\chi_{jk}^{EB}(-\omega,\bm{B}),~\kappa_{ijk}^{BBE}(\omega)=\frac{1}{2}\lim_{\bm{B\to0}}\partial_{B_i}\chi_{jk}^{BE}(-\omega,\bm{B}).
                \end{equation}
Here we defined the magnetoelectric susceptibility 
                \begin{equation}
                P_i (\omega) = \chi_{ij}^{EB} (\omega) B_j (\omega),~M_i (\omega) = \chi_{ij}^{BE} (\omega) E_j (\omega)
                \label{optical_magnetoelectric_effect}
                \end{equation}
representing the odd-parity coupling between electric and magnetic polarizations such as magnetoelectric effect and (magnetic-dipole-related) natural optical activity~\cite{Dzyaloshinskii1960-ad,Astrov1960-lc,Rado1961-oq}.
Specifically, the Pitaevskii relation of Eq,~\eqref{EH_to_magnetization_pitaevskii} has been partly elaborated in studies of isotropic and nonmagnetic media, for which the rectification response is termed with the inverse magnetochiral effect~\cite{Wagniere1989-xm,Wozniak1991-ww,Wozniak1992-uq}.
Equation~\eqref{EH_to_magnetization_pitaevskii} is the generalization of the inverse magnetochiral effect and is thus applicable to various cases such as the \PT{}-symmetric materials and anisotropic media.
Furthermore, \T{}-\PT{} classification allows us to take a closer look at the response as follows.

By using the \T{} and \PT{} parities given by $\tau_{\theta}^\text{tot} = +1, \tau_{\theta I}^\text{tot} = -1$, we can decompose RMI into those allowed in the \T{}-symmetric and \PT{}-symmetric materials.
In the \T{}-symmetric case, following the parallel discussion on the inverse Faraday effect, only the symmetric part of the magnetoelectric susceptibility ($\chi_{jk}^{CC} = \chi_{kj}^{CC} $) has a nonvanishing derivative with respect to the magnetic field $\bm{B}$.
Then, according to the Pitaevskii relation of Eq.~\eqref{EH_to_magnetization_pitaevskii}, the formula for RMI satisfies
                \begin{equation}
                        \Theta_{ijk} = \Theta_{ikj},~ \Imag{\Theta_{ikj}} =0,
                        \label{reciprocal_magnetization_symmetric}
                \end{equation}
in the \T{}-symmetric materials.
Since the relevant magnetoelectric susceptibility ($\Real{\chi_{jk}^{CC}}$) represents the optical magnetoelectric effect, which is the AC analog of the (static) magnetoelectric effect and is \T{}-odd and \PT{}-even~\cite{Malashevich2010-uw}.
Then, we call the rectification response denoted by $\Real{\Theta_{ijk}} $ the \textit{inverse (optical) magnetoelectric effect} in the same spirit of the inverse Faraday effect.
In the case of isotropic and nonmagnetic media, the tensor of rectification response is reduced to $\Theta_{ijk} = \Theta_0 \, \epsilon_{ijk}$ ($\Theta_0 \in \mathbb{R}$).
Obtained response functions $\Theta_0$ represent the inverse magnetochiral effect~\cite{Wozniak1991-ww}, a specific case of the inverse magnetoelectric effect.

Importantly, owing to the \PT{}-ensured Onsager reciprocity of Eq.~\eqref{Onsager-relation_general-case_PT-parity}, the components in $\Theta_{ijk}$ for the inverse magnetoelectric effect vanishes in the \PT{} symmetric system.
On the other hand, the \T{} violation allows for non-zero $\bm{B}$ derivative for the antisymmetric part of the magnetoelectric susceptibility ($\chi_{jk}^{CC} = - \chi_{kj}^{CC} $) as in the case of inverse Cotton-Mouton effect.
Thus, if the system does not respect the \T{} symmetry but \PT{} symmetry, only the antisymmetric part contributes to RMI, resulting in the relation
                \begin{equation}
                        \Theta_{ijk} = -\Theta_{ikj},~ \Real{ \Theta_{ikj}} =0,  
                        \label{reciprocal_magnetization_antisymmetric} 
                \end{equation}
in \PT{}-symmetric systems.
Such a \T{}-violating but \PT{}-symmetric system can be found in a series of antiferromagnets~\cite{Fiebig2005-hj}.
The anti-symmetric part ($\Imag{ \chi_{jk}^{CC}}$) means the natural optical activity arising from the correlation between the electric and magnetic dipole transitions, which is observed in nonmagnetic and \Pa{}-violating materials such as ferroelectric materials.
Then, the rectification response of Eq.~\eqref{reciprocal_magnetization_antisymmetric} is the \textit{inverse natural optical activity} characteristic of odd-parity and \PT{}-symmetric magnetic materials.

One may obtain an intuitive picture of the field-induced natural optical activity as follows.
The \PT{}-symmetric magnetic order gives rise to the coupling between the magnetic field and the noncentrosymmetric and \T{}-symmetric fields; \textit{e.g.}, the magnetoelectric coupling is the linear coupling between the magnetic and electric fields [DC limit of Eq.~\eqref{optical_magnetoelectric_effect}].
The interplay between the \PT{}-symmetric magnetic and $\bm{B}$ fields leads to the nonmagnetic \Pa{} violation and thereby realizes the natural optical activity under $\bm{B}$.
To summarize the space-time classification, RMI is determined by the inverse magnetoelectric effect in the presence of \T{} symmetry, while it is by the inverse natural optical activity in the \PT{}-symmetric systems.

Finally, let us consider the polarization state of light relevant to RMI.
The tensor symmetry and property of complex number [Eqs.~\eqref{reciprocal_magnetization_symmetric}~\eqref{reciprocal_magnetization_antisymmetric}] indicate that the inverse magnetoelectric and inverse natural optical activity are the responses to the $\Real{ E_j^\ast B_k }$ and $\Imag{ E_j^\ast B_k }$, respectively.
When the electromagnetic field satisfies $\bm{B} =  \bk \times \bm{E}/ \omega$ with the wave vector $\bk$ of light as it does in the vacuum, the double external fields are recast as
        \begin{equation}
        E_j^\ast B_k = \frac{1}{\omega} \epsilon_{k\alpha\beta } k_\alpha ~  E_j^\ast E_\beta. 
        \end{equation}
Thus, similarly to the nonreciprocal magnetization induction such as the inverse Faraday effect, the imaginary part is present if the light includes the circular component, while the real part is non-zero in general due to $|\bm{E}|^2$. 
It follows that the inverse magnetoelectric effect occurs even when the light is not circularly-polarized, while the inverse natural optical activity does under the circularly-polarized-light irradiation.

Note that the electric-quadrupole field $E^Q_{ab} \equiv \left( \partial_a E_b + \partial_b E_a \right)/2$ gives electromagnetic excitations comparable to that from the magnetic-dipole field ($\bm{B}$) in the gradient expansion of the electromagnetic field.
Then, up to the lowest-order contributions including RMI, the formula for the DC magnetization response is 
                \begin{equation}
                M_i = \kappa_{ijk}^{BEE} E_j^\ast E_k + \kappa_{ijk}^{BEB} E_j^\ast B_k + \kappa_{ijk}^{BBE} B_j^\ast E_k + \kappa_{ij(kl)}^{BEQ} E_j^\ast E^Q_{kl} + \kappa_{i(jk)l}^{BQE} (E^Q_{jk})^\ast E_l, 
                \label{RMI_withEQ}
                \end{equation}
where we introduced the odd-parity DC magnetization response $\kappa_{ij(kl)}^{BEQ}, \kappa_{i(jl)k}^{BQE}$ to the electric-dipole ($E$) and electric-quadrupole ($Q$) fields.
Since the total \T{} parity of $\kappa_{ij(kl)}^{BEQ}$ is opposite to $\Theta_{ijk}$, $\Imag{ \kappa_{ij(kl)}^{BEQ}} $ ($\Real{ \kappa_{ij(kl)}^{BEQ}}$) contributes to RMI in the \T{}-symmetric (\PT{}-symmetric) systems in contrast to Eq.~\eqref{reciprocal_magnetization_symmetric} [Eq.~\eqref{reciprocal_magnetization_antisymmetric}].
Note that the linear responses relevant to the rectification responses $\kappa_{ij(kl)}^{BEQ}$ share the same symmetry with the piezoelectric effect and magnetopiezoelectric effects allowed in the \T{}- and \PT{}-symmetric materials, respectively~\cite{Varjas2016-sw,Watanabe2017-qk,Shiomi2019-co}.

In isotropic media, the response formula of Eq.~\eqref{RMI_withEQ} is recast as
                \begin{equation}
                \bm{M} = i\kappa_0  \bm{E}^\ast \times \bm{E}  + 2\Real{\Theta_0 \left( \bm{B}^\ast \times \bm{E} \right) + \Gamma_1  \nabla | \bm{E} |^2 + \Gamma_2  \left( \nabla \cdot \bm{E}^\ast \right)  \bm{E} + \Gamma_3  \left( \nabla \cdot \bm{E} \right)  \bm{E}^\ast },
                \end{equation}
where we implicitly assume either \T{} or \PT{} symmetry by which the inverse Cotton-Mouton effect is forbidden.
Since $\bm{B}^\ast \times \bm{E}$ is real and $\nabla \cdot \bm{E} = 0$ if the monochromatic-field conditions such as $\bm{B} = \bk \times \bm{E} / \omega$ hold, RMI is attributed to $\Real{ \Theta_0}$ and $\Real{ \Gamma_1}$, which are allowed in the \T{}-symmetric and \PT{}-symmetric systems, respectively.

\subsection{Tabulation of Pitaevskii relations}
\label{SecSub_symmetry_tabulation}

The generalized Pitaevskii relations of Eq.~\eqref{generalized_Pitaevskii-relation} allow us to predict connections between the rectification and linear responses.
Let us consider examples by taking auxiliary fields as $\bm{F} = \bm{E}, \bm{B}, \hat{\sigma}$ where $\hat{\sigma}$ is the stress conjugate to the strain $\hat{\varepsilon}$. 
We can obtain 18 Pitaevskii relations in total from the 6 linear susceptibility tensors
                \begin{equation}
                \chi_{jk}^{EE},\chi_{jk}^{BB},\chi_{(jk)(lm)}^{\sigma\sigma},\chi_{jk}^{EB},\chi_{jk}^{B\sigma},\chi_{(jk)l}^{\sigma E}, 
                \end{equation}
undergoing the correction proportional to $\bm{F} = \bm{E}, \bm{B}, \hat{\sigma}$.
The linear-susceptibility tensors are for the electric permittivity ($\chi_{jk}^{EE}$), magnetic permittivity ($\chi_{jk}^{BB}$), elastic susceptibility ($\chi_{(jk)(lm)}^{\sigma\sigma}$), magnetoelectric susceptibility ($\chi_{jk}^{EB}$), piezomagnetic susceptibility ($\chi_{jk}^{B\sigma}$), and piezoelectric susceptibility ($\chi_{(jk)l}^{\sigma E}$).
The Pitaevskii relations are tabulated in Table~\ref{Table_pitaevskii}.
It suffices to show the results related to $\chi_{jk}^{EE}$ and $\chi_{jk}^{EB}$, manifesting the opposite \T{} parity.
The Pitaevskii relations concerning other linear responses are straightforwardly obtained.

                \begin{table}[htbp]
                \caption{
                        Relations between the linear and rectification responses relevant to the electric field $\bm{E}$, magnetic field $\bm{B}$, and stress $\hat{\sigma}$.
                        Bearing the rectification response $X_i = \kappa_{ijk} F_j^\ast F_k$ in mind, `Linear response' is defined for the physical fields $(F_j, F_k)$ and $F_i$ is conjugate to the rectified response $X_i$.
                        $\tau_\theta^\text{tot} = \pm 1$ denotes the time-reversal parity of $\kappa_{ijk}$.
                        Real and imaginary parts of $\hat{\kappa}$ are classified by whether it is allowed without or with the \T{} violation (see also Table~\ref{Table_classification_pitaevkii_T_symmetry}).
                        `Rectification' is for the rectification response and available references.
                        Some entries in `$F_i$' have the superscript `$\ddagger$' to denote the \Pa{}-odd parity of the corresponding rectification responses, and therefore either \T{}-even or \T{}-odd contribution is forbidden if the \PT{} symmetry is preserved.
                        }
\label{Table_pitaevskii}
                \centering
                \begin{tabular}{cCCCcccc}
                 \toprule
                 Linear response&(F_j,F_k)&F_i&\tau_\theta^\text{tot}&Re/Im\,[$\kappa_{ijk}$]&\T{}-even&\T{}-odd&Rectification\\
                 \midrule
                 \midrule
                 electric susceptibility&(\bm{E},\bm{E})&\bm{E}^\ddagger&+1&Re &\cm&&Optical rectification~\cite{Bass1962-lf} \\
                 &&&&Im &&\cm&Optical magneto-rectification \\
                 &&\bm{B}&-1&Re &&\cm&Inv. Cotton-Mouton~\cite{Kalashnikova2007-yx,Hansteen2006-zg} \\
                 &&&&Im &\cm&&Inv. Faraday~\cite{Pitaevskii1961-rd,van-der-Ziel1965-dm} \\
                 &&\hat{\sigma}&+1&Re &\cm&& Optical electrostrictive \\
                 &&&&Im &&\cm& Optical magneto-electrostrictive\\
                 \midrule
                 magnetoelectric susceptibility&(\bm{E},\bm{B})&\bm{E}&-1&Re &&\cm&Inv. magneto-electrogyration \\
                 &&&&Im &\cm&&Inv. electrogyration \\
                 &&\bm{B}^\ddagger&+1&Re &\cm&&Inv. magnetoelectric~\cite{Wagniere1989-xm} \\
                 &&&&Im &&\cm&Inv. natural optical activity\\
                 &&\hat{\sigma}^\ddagger&-1&Re &&\cm& Optical piezomagnetoelectric \\
                 &&&&Im &\cm&& Kinetic piezomagnetoelectric\\
                 \bottomrule
                \end{tabular}
                \end{table}

For instance, our classification identifies the inverse phenomenon of the electrogyration effect associated with $(F_i, F_j, F_k) = (\bm{E},\bm{E},\bm{B})$.
The electrogyration effect, the $\bm{E}$-induced optical activity, has been demonstrated in theory and experiment~\cite{Aizu1964-rr,Zheludev1964,Vlokh1970,Prosandeev2013-da} and applied to the imaging of the \Pa{}-even symmetry breaking effect such as ferroaxial order~\cite{Hayashida2020-tu}~\footnote{
    The inverse electrogyration effect can be regarded as another inverse phenomenon of the natural optical activity, concerning the stress $\hat{\sigma}$ instead of the magnetic field $\bm{B}$ in the case of the ``inverse natural optical activity'' in Table~\ref{Table_pitaevskii}.
    To highlight this difference, we can also call the inverse electrogyration, \textit{i.e.}, the rectification responses of the strain connected with $\Imag{\chi_{ij}^{EB}(\omega)} $, the inverse $\hat{\sigma}$-induced natural optical activity.
    Similarly, ``inverse natural optical activity'' in Table~\ref{Table_pitaevskii} should be understood as the inverse $\bm{B}$-induced natural optical activity. 
    Similar things can also be said for the other responses in Table~\ref{Table_pitaevskii}.
}.
Note that one should treat the other electrogyration effect related to the electric-quadrupole excitations on equal footing.
The corresponding rectification response is 
                \begin{equation}
                P_a = \kappa_{ij(kl)}^{EEQ} (\omega) E_j^\ast (\omega) E_{kl}^Q (\omega),
                \end{equation}
whose response function is related with $\partial_{E_i} \chi_{j(kl)}^{EQ}$ via the Pitaevskii relation.
The response is similar to the so-called electric-quadrupole second-harmonic generation~\cite{Fiebig2005-ne,Jin2020-xk}.

For responses including the elastic degree of freedom in Table~\ref{Table_pitaevskii}, let us consider the photo-induced strain response given by
                \begin{equation}
                \varepsilon_{ij} = \kappa_{(ij)kl}^{\sigma EE} (\omega) E_k^\ast (\omega) E_l(\omega).
                \label{EE_to_strain_rectification}
                \end{equation} 
In the DC limit, the coupling between $\hat{\varepsilon}$ and $E_k E_l$ indicates the electrostrictive effect.
Thus, the Pitaevskii relation claims that the optical electrostrictive effect we defined by Eq.~\eqref{EE_to_strain_rectification} is related to the electric susceptibility modified by the stress.

One also notice an odd-parity strain response 
                \begin{equation}
                \varepsilon_{ij} = \kappa_{(ij)kl}^{\sigma EB} (\omega) E_k^\ast (\omega) B_l(\omega),
                \label{EB_to_strain_rectification}
                \end{equation} 
which is correlated with the stress-induced optical magnetoelectric coupling according to the Pitaevskii relation.
In the DC limit, the response indicates the trilinear coupling between the strain, electric polarization, and magnetization, namely piezomagnetoelectric effect~\cite{Rado1962-av,Fil2017-ns}.

Note that one can further exploit the Pitaevskii relations by taking another auxiliary field such as the spin gauge fields for the spin-current response, sublattice-dependent magnetic field~\cite{Merte2023-sk}, the spatial gradient of the phase of the superconducting order parameter.
For instance, the nonreciprocal current generation in the superconducting phase is related to the non-absorptive linear optical conductivity~\cite{Watanabe2022-hk}.

\section{Numerical study of reciprocal magnetization induction}
\label{Sec_reciprocal_magnetization_induction}

In this section, we verify that our formulation of the generalized Pitaevskii relation is in agreement with the numerical results.
For a specific case, let us consider RMI, comprised of the inverse optical magnetoelectric effect and inverse natural optical activity [Eq.~\eqref{EH_to_magnetization}].
To corroborate the contrasting space-time symmetry of the two effects, we adopt toy models where the \Pa{} symmetry is violated while either \T{} or \PT{} symmetry is retained.
We adopt the normalized lattice constant and the natural units such as $\hbar =1$ for the Dirac constant and $e=1$ for the elementary charge of fermion.

\subsection{Setup}
\label{SecSub_setup}

\begin{figure}[htbp]
                \centering
                \includegraphics[width=0.40\linewidth,clip]{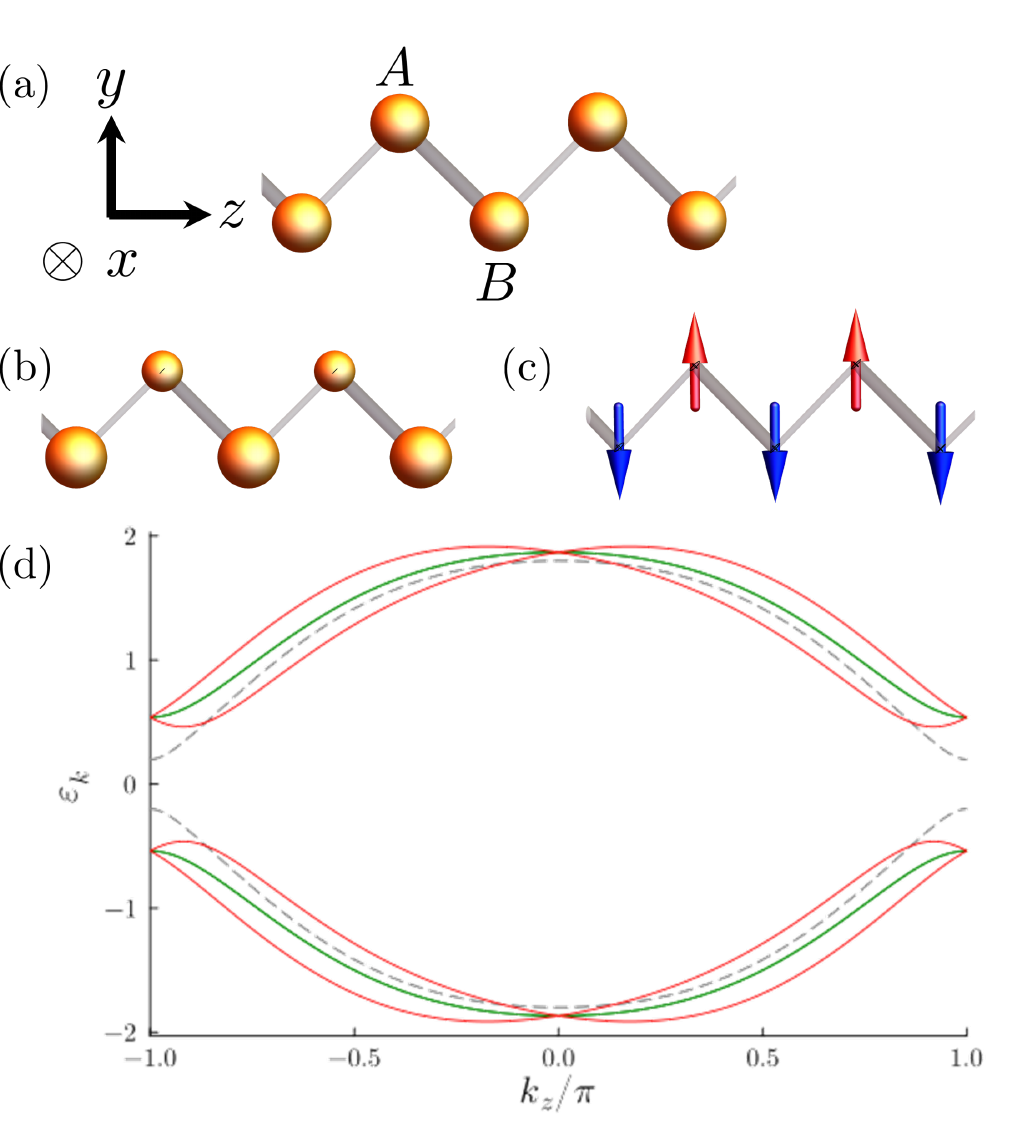}
                \caption{
                        (a) zigzag chain comprised of $A$ and $B$ sublattices.
                        Dimerization is denoted by the thick lines.
                        (b) \T{}-symmetric and \Pa{}-broken state resulting from the staggered onsite potential.
                        (c) \PT{}-symmetric and \Pa{}-broken state due to the antiferromagnetic order.
                        (d) Band structures of the para-state (dashed line), that in the case of panel (b) (red solid line), and that in the case of panel (c) (green solid line). 
                        }
                \label{Fig_zigzag}
                \end{figure}

The model is one-body tight-binding Hamiltonian for the one-dimensional zigzag chain comprised of two sublattice $(A,B)$ [Fig.~\ref{Fig_zigzag}(a)].
The Hamiltonian of the spinful fermions reads as
                \begin{equation}
                \mathcal{H} = \sum_{\bk} \bm{c}_\bk^\dagger \mathrm{H}_\bk \bm{c}_\bk,
                \end{equation}
where $\bm{c}_\bk = \left(  c_{\bk A \uparrow},c_{\bk A \downarrow},c_{\bk B \uparrow}, c_{\bk B \downarrow} \right)^T$ is the vector of annihilation operators for the fermion labeled by the crystal momentum $\bk$, sublattice ($\tau = A,B$), and spin ($\uparrow,\downarrow$).
Then, the many-body energy eigenstates are spanned by the Fock space with the one-body energy eigenstates.
The occupation is given by the Fermi-Dirac distribution function $f_a^\text{FD} = (\exp{((\epsilon_a-\mu))/T}+1)^{-1}$ parametrized by temperature $T$ and the chemical potential $\mu$.

The Bloch Hamiltonian $\mathrm{H}_\bk$ consists of the centrosymmetric ($\mathrm{H}_0 (\bk)$) and \Pa{}-violating ($\mathrm{H}_1 (\bk)$) parts.
The centrosymmetric term is given by
                \begin{equation}
                \mathrm{H}_0 (\bk) = - (t+u)\cos{\frac{k_z}{2}} \tau_x- (t-u)\sin{\frac{k_z}{2}} \tau_y + \lambda \sin{k_z} \sigma_x \tau_z,
                \end{equation}
and satisfies the $\mathcal{P}$ symmetry $\tau_x\mathrm{H}_0(-\bm{k})\tau_x=\mathrm{H}_0(\bm{k})$.
The pauli matrices $\bm{\sigma}$ and $\bm{\tau}$ are for the spin and sublattice degrees of freedom.
$t = 1,u=0.8$ are the nearest-neighbor hoppings, by which $t-u$ denotes the dimerization between neighboring sites, and $\lambda=0.6$ is the sublattice-dependent spin-orbit coupling~\cite{Kane2005-lv,Zelezny2014-qr,Yanase2014-mw}.

Let us take into account the noncentrosymmetric term $\mathrm{H}_1 (\bk)$ in the following two-fold manners.
In the \T{}-symmetric case, the parity-breaking effect is given by the staggered potential like the Su–Schrieffer–Heeger model
                \begin{equation}
                        \mathrm{H}_{1} (\bk) = \mathrm{H}_{\theta} (\bk) = \delta \tau_z, 
                \end{equation}
which breaks the symmetry about the \Pa{} operation defined with the A-B bond center [Fig.~\ref{Fig_zigzag}(b)].
The symmetry-breaking effect induces the spin-momentum splitting with preserving the degeneracy between $\pm \bk$ protected by the \T{} symmetry [Fig.~\ref{Fig_zigzag}(d)].
On the other hand, the \PT{}-even but \Pa{}-breaking effect is built into the Hamiltonian by 
                \begin{equation}
                        \mathrm{H}_{1} (\bk) = \mathrm{H}_{\theta I } (\bk) = h_0 \sigma_y \tau_z. 
                \end{equation}
$h_0$ is the molecular field of the antiferromagnetic ordering.
The magnetic moments are aligned to the $y$-axis and staggered between the $A$ and $B$ sites [Fig.~\ref{Fig_zigzag}(c)].
The energy spectrum remains doubly degenerate at each crystal momentum due to the \PT{} symmetry  [Fig.~\ref{Fig_zigzag}(d)].
In the following, we demonstrate RMI and its Pitaevskii relations based on the \T{}-symmetric Hamiltonian $\mathrm{H}_0 + \mathrm{H}_\theta $ and the \PT{}-symmetric one $\mathrm{H}_0 + \mathrm{H}_{\theta I } $ with $\delta,\,h_0 =0.5$.

The physical responses are numerically calculated with the formulas of Eqs.~\eqref{RDM_full_2nd_optical_conductivity},~\eqref{linear_susceptibility_function}.
The physical fields are $(F_i,F_j,F_k) = (\bm{B},\bm{B},\bm{E})$ for the reciprocal magnetization induction of Eq.~\eqref{EH_to_magnetization} and $(F_i,F_j) = (\bm{B},\bm{E})$ for the magnetoelectric susceptibility of Eq.~\eqref{optical_magnetoelectric_effect}.
The electric field is expressed in the velocity gauge with which the photo-electric field is $\bm{E} = i \omega \bm{A}$.
Each physical field is coupled to the fermions as 
                \begin{equation}
                \mathrm{H}_\text{B} (\bk) = \bm{B} \cdot \bm{\sigma}, 
                \end{equation}
for the magnetic field $\bm{B}$ and 
                \begin{equation}
                \mathrm{H}_\bk \to \mathrm{H}_{\bk + \bm{A}},
                \end{equation}
for the vector potential $\bm{A}$.
After calculating the correlation functions $\chi_{ij}^{BA}$ and $\kappa_{ijk}^{BBA}$, we obtain the response functions $\chi_{ij}^{BE}$ and $\kappa_{ijk}^{BBE}$ of interest by using $\bm{E} = i \omega \bm{A}$.
We replace the adiabaticity parameter $\eta =+0$ with the phenomenological scattering rate $\gamma >0$.
We adopt $T=10^{-3}$ and $\gamma=10^{-3}$ unless explicitly mentioned.
For the $\bk$ integration, we adopt the $N$-discretized first Brillouin zone ($N=10^4$).

\subsection{Reciprocal magnetization induction and Pitaevskii relations}
\label{SecSub_RMI_numerical}

Let us consider the symmetry of the adopted Hamiltonians and the allowed responses.
The \T{}-symmetric model is labeled by the magnetic point group
                \begin{equation}
                m1',
                \end{equation}
with the $yz$ mirror symmetry, while the \PT{}-symmetric model is by
                \begin{equation}
                2/m',
                \end{equation}
with the two-fold rotation along the $x$-axis.
The noncentrosymmetric symmetry allows for the $\bm{B}$-linear correction to the magnetoelectric susceptibility ($\partial_{B_i} \chi_{jz}^{BE}$)
                \begin{equation}
                (ij) = (xx), (yy), (zz),(yz),(zy),
                \end{equation}
in the \T{}-symmetric Hamiltonian and 
                \begin{equation}
                (ij) = (xy), (yx), (zx),(xz),
                \end{equation}
for the \PT{}-symmetric case.
Note that we consider the electric field along the $z$ direction ($\bm{E}\parallel \hat{z}$) due to the one-dimensional Hamiltonian.
Since the tensor shapes of the $\kappa_{ijk}^{BBE}$ and $\partial_{B_i} \chi_{jz}^{BE}$ coincide with each other, the allowed components of RMI are obtained in parallel.
We corroborate $\kappa_{yyz}^{BBE}$ and $\kappa_{yxz}^{BBE}$ of \T{}- and \PT{}-symmetric systems respectively, though the qualitative aspects do not change for the other components.

\begin{figure}[htbp]
                \centering
                \includegraphics[width=0.50\linewidth,clip]{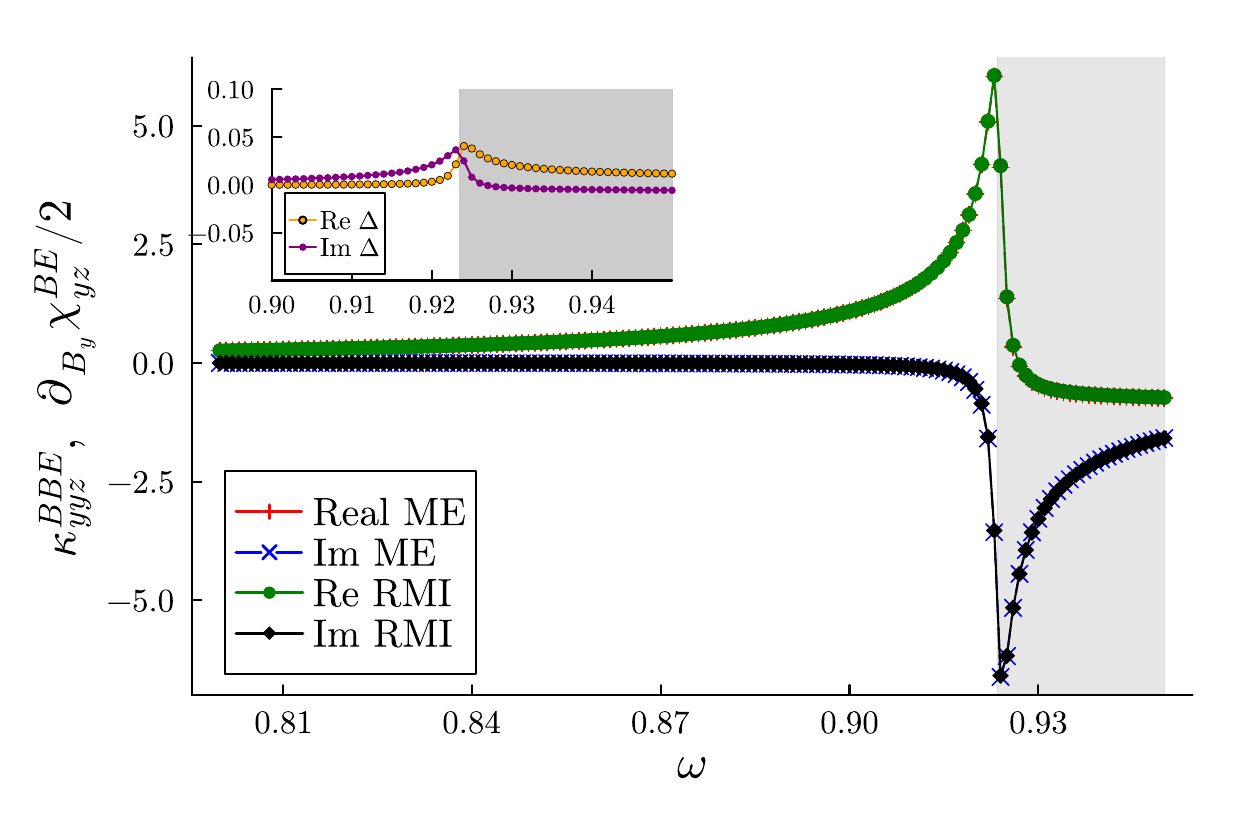}
                \caption{
                        Spectrum of the reciprocal magnetization induction $\kappa_{yyz}^{BBE} (\omega)$ (RMI) and magnetoelectric susceptibility $\partial_{B_y} \kappa_{yz}^{BE}(\omega)$ (ME) of the \T{}-symmetric model.
                        The resonant particle-hole excitations are present in the shaded area.  
                        The chemical potential is $\mu=0$ corresponding to the insulator phase.
                        Real (Imaginary) parts of two response functions almost overlap with each other. 
                        (inset) Spectrum of $\Delta = \partial_{B_y} \kappa_{yz}^{BE} - \kappa_{yyz}^{BBE}$.
                        The deviation gets negligible well below the optical gap.
                        }
                \label{Fig_RMI_tsymmetric}
                \end{figure}

First, let us consider the \T{}-symmetric case.
The spectrum of RMI and $\bm{B}$-modified magnetoelectric susceptibility is shown in Fig.~\ref{Fig_RMI_tsymmetric}.
The chemical potential is set to $\mu=0$, and the system is in the band-insulator phase.
In accordance with Eqs.~\eqref{reciprocal_magnetization_symmetric}, each response function is real below the optical gap ($\omega \leq 0.924$), while the imaginary parts also participate in responses since resonant particle-hole excitations break down the non-absorption condition.
Note that we show the full susceptibility $\kappa$ including the absorptive part $\kappa^\text{a}$ as well as $\kappa^\text{na}$.
It is evident from the in-gap spectrum shown in Fig.~\ref{Fig_RMI_tsymmetric} that the Pitaevskii relation for the inverse magnetoelectric effect holds.
Although $\kappa_{yyz}^{BBE}$ almost coincides with $\partial_{B_y} \kappa_{yz}^{BE}$ in the entire frequency range in Fig.~\ref{Fig_RMI_tsymmetric} due to the simplicity of the model Hamiltonian, the deviation $\Delta \equiv \partial_{B_y} \kappa_{yz}^{BE} - \kappa_{yyz}^{BBE}$ develops around the optical gap when the frequency of light increases as emphasized in the inset.
The real part of the deviation Re~$\Delta$ is vanishingly small below the optical gap (within the ambiguity of the order $T,\gamma=10^{-3}$) as expected from the Pitaevskii relation Re~$\Delta=0$, while it takes finite values above the optical gap.
We also checked that Im~$\Delta$ vanish well below the optical gap as both Im~$\kappa_{yyz}^{BBE}$ and $\partial_{B_y} \Imag{\chi_{B_yE_z}}$ should vanish there.

\begin{figure}[htbp]
                \centering
                \includegraphics[width=0.50\linewidth,clip]{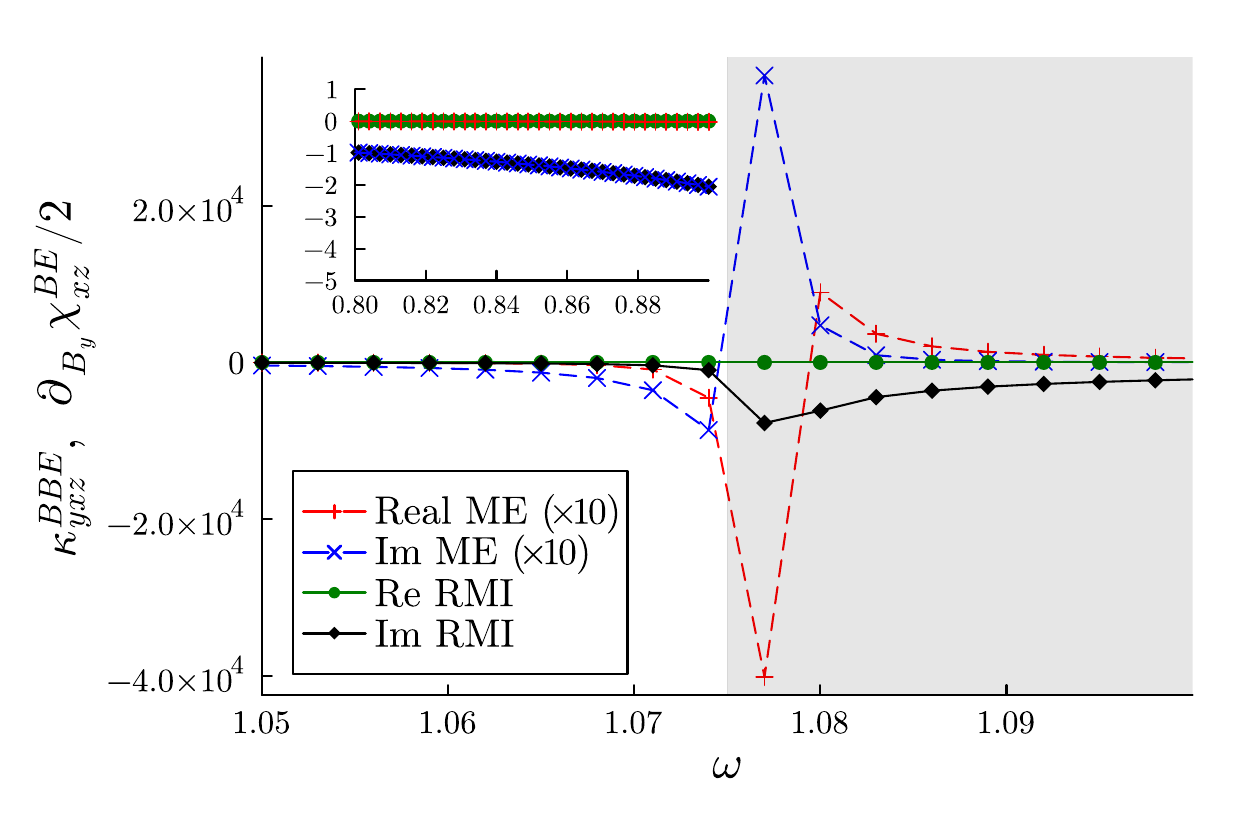}
                \caption{
                        Spectrum of the reciprocal magnetization induction $\kappa_{yxz}^{BBE} (\omega)$ (RMI) and magnetoelectric susceptibility $\partial_{B_y} \kappa_{xz}^{BE} (\omega)$ (ME) of the \PT{}-symmetric model.
                        The resonant particle-hole excitations are present in the shaded area.  
                        The chemical potential is $\mu=0$ corresponding to the insulator phase.
                        (inset) Enlarged view of the spectrum in the in-gap regime.
                        Note that no multiplication is applied to each response plotted in the inset, different from the main plot.
                        The imaginary parts of two responses coincide with each other, whereas the real parts are vanishingly small.
                        }
                \label{Fig_RMI_ptsymmetric}
                \end{figure}

Next, we investigate the \PT{}-symmetric case.
Assuming the band-insulator state with $\mu=0$, we obtain the frequency spectrum of responses as shown in Fig.~\ref{Fig_RMI_ptsymmetric}.
In contrast to the \T{}-symmetric case, one can observe good coincidence between $\partial_{B_y} \Imag{ \kappa_{yz}^{BE}(\omega)}$ and $\Imag{ \kappa_{yxz}^{BBE} (\omega)}$ below the optical gap (inset of Fig.~\ref{Fig_RMI_ptsymmetric}).
It follows that the Pitaevskii relation holds for the inverse natural optical activity.

The numerical evidence supports the validity of the Pitaevskii relations for the inverse optical magnetoelectric effect and natural optical activity of Eq.~\eqref{EH_to_magnetization_pitaevskii}.
The relation holds if and only if the non-absorption condition is satisfied and the distribution-modulation effect is negligible, and thus one may be interested in how the Pitaevskii relation ceases to be satisfied.
We have already observed that the relations do not hold if the light irradiation allows for electron-hole excitations (Figs.~\ref{Fig_RMI_tsymmetric},~\ref{Fig_RMI_ptsymmetric}).  
Then, let us consider the effect of the distribution-modulation effect [Eq.~\eqref{rectification_fermi-surface_factor}]. 

\begin{figure}[htbp]
                \centering
                \includegraphics[width=0.50\linewidth,clip]{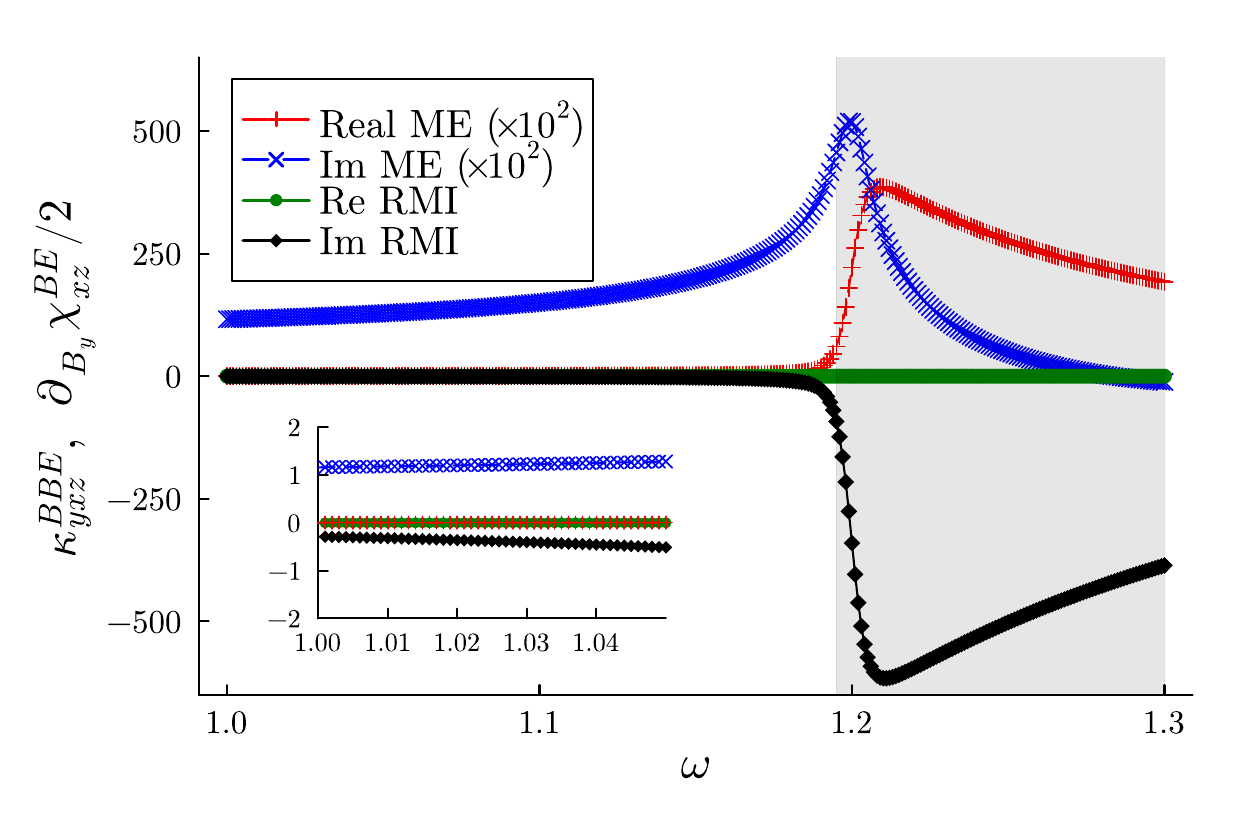}
                \caption{
                        Same plots as those in Figure~\ref{Fig_RMI_ptsymmetric}, while the chemical potential is $\mu=-0.6$ corresponding to the metal phase.
                        We note that the optical gap is shifted to $\omega \sim 1.2$, different from that of Fig.~\ref{Fig_RMI_ptsymmetric}. 
                        (inset) Enlarged view of the spectrum in the in-gap regime.
                        }
                \label{Fig_RMI_ptsymmetric_with_Fermi_surface}
                \end{figure}

In the adopted Hamiltonian, the distribution modulation can occur in the presence of the Fermi surface.
Here we focus on the \PT{}-symmetric case and take $\mu=-0.6$.
The metallic conductivity is identified by the low-frequency spectrum of the optical conductivity (not shown).
Figure~\ref{Fig_RMI_ptsymmetric_with_Fermi_surface} shows the spectrum of RMI and the $\bm{B}$-derivative of magnetoelectric susceptibility around the optical gap.
Both of $\partial_{B_y} \Imag{ \kappa_{yz}^{BE}(\omega)}$ and $\Imag{ \kappa_{yxz}^{BBE} (\omega)}$ are non-zero below the optical gap as well (inset of Fig.~\ref{Fig_RMI_ptsymmetric_with_Fermi_surface}), whereas they show the significant deviation and thereby indicates the breakdown of the Pitaevskii relation.

\begin{figure}[htbp]
                \centering
                \includegraphics[width=0.5\linewidth,clip]{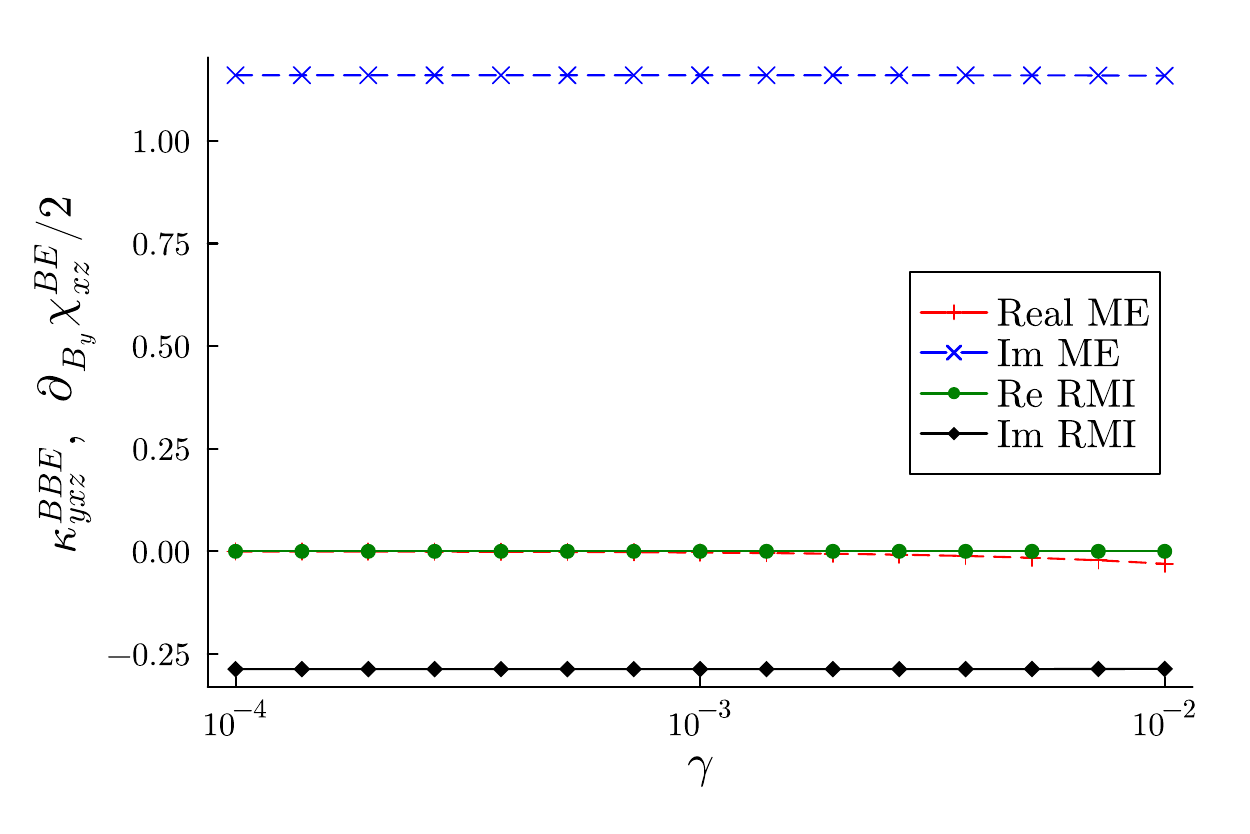}
                \caption{
                        Phenomenological-scattering-rate dependence of the reciprocal magnetization induction $\kappa_{yxz}^{BBE} (\omega)$ (RMI) and magnetoelectric susceptibility $\partial_{B_y} \kappa_{xz}^{BE} (\omega)$ (ME) of the \PT{}-symmetric model.
                        The frequency of light is $\omega_0 = 0.5$, and the chemical potential is $\mu=-0.6$ where the metallic conductivity is present.
                        }
                \label{Fig_RMI_ptsymmetric_with_Fermi_surface_scattering_effect}
                \end{figure}

The obtained deviation is not attributed to the smearing of the resonant contributions.
To eliminate that possible extrinsic effect, we calculate RMI and the magnetoelectric susceptibility with varying the phenomenological scattering rate $\gamma$ (Fig.~\ref{Fig_RMI_ptsymmetric_with_Fermi_surface_scattering_effect}).
The frequency is fixed to that below the optical gap as $\omega_0 = 0.5$.
Despite the increasing scattering rate over the orders as $\gamma = 10^{-4} \sim 10^{-2}$, $\partial_{B_y} \Imag{ \kappa_{yz}^{BE}(\omega)}$ and $\Imag{ \kappa_{yxz}^{BBE} (\omega)}$ do show negligible variation, indicating that the difference between them is the intrinsic behavior free from the smearing effect.
This is also evident from the comparison between $\partial_{B_y} \Real{ \kappa_{yz}^{BE}(\omega)}$ and $\partial_{B_y} \Imag{ \kappa_{yz}^{BE}(\omega)}$, former of which undergoes slight modification by the increasing scattering effect.

As a result, the insulating state is necessary in addition to the non-absorption condition for the validity of the Pitaevskii relations.
The distribution-modulation effect is similarly observed in the system comprised of multiple degrees of freedom such as electron and phonon.
In those systems, the distribution may be modified under the external stimuli with the non-absorption condition kept.
It is an important future work to further elucidate the possible breakdown of the Pitaevskii relation without phenomenological treatments of scattering effects.

\section{Summary}
\label{Sec_summary}

We have formulated the rectification and Pockels effects satisfying the non-absorption condition and generalized Pitaevskii's argument to cover various correlations between the linear and rectification responses.
In a full-quantum manner, we obtained the Pitaevskii relation by which the linear, rectification, and Pockels responses are closely related to each other if there is neither interband-like nor intraband-like excitation.
The derivation based on many-body energy eigenstates does not depend on any specific approximation and is thus applicable to various systems.
Although the Pitaevskii relations have been investigated in previous works in terms of the inverse magneto-optical responses and optical rectification, the derived generalized Pitaevskii relations are applicable to diverse physical phenomena such as cross-correlation between electric, magnetic, elastic, and other degrees of freedom.
For instance, we identified a series of Pitaevskii relations and systematically classified them in terms of space-time symmetry (Table~\ref{Table_pitaevskii}).
The analytical results are supported by numerical calculations of the inverse magnetoelectric effect and inverse natural optical activity.
The numerical results further imply that the Pitaevskii relation may be violated in the presence of the Fermi-surface effect even when the frequency of light is in the off-resonant regime.
These analytical and numerical demonstrations of the Pitaevskii relations may offer implications for future studies utilizing the nonlinear light-matter coupling and thereby spark further interest in ultrafast spintronic phenomena.

\section*{Acknowledgement}

H.W. is grateful to Masakazu Matsubara for his fruitful comments.
The authors are supported by JSPS KAKENHI
No.~JP23K13058 (H.W.),
No.~JP21K13880,	No.~JP22H04476, No.~JP23K17353, No.~JP24H01662 (A.D.).

\appendix

\section{Derivation of off-resonant rectification response}
\label{SecApp_derivation_off-resonant_rectification}

We derive the non-absorptive rectification response function from Eq.~\eqref{RDM_full_2nd_optical_conductivity}.
The general formula is given by  
        \begin{align*}
        X^i_{(2)} (\omega)
             &= \int \frac{d\omega_1 d\omega_2 }{(2\pi)^2} 2\pi \delta (\omega -\omega_1 - \omega_2) F_{j}(\omega_1) F_{k} (\omega_2) \notag  \\
             &\times \lim_{\bm{F}\to \bm{0}} \Biggl[ \sum_a \frac{1}{2}  \vj^{ijk}_{aa} f_a  \\ 
             &+\sum_{a,b} \frac{1}{2} \frac{\vj^{ij}_{ab} \vj^k_{ba} f_{ab} }{ \omega_2 +i\eta -\epsilon_{ba}} + \frac{1}{2} \frac{\vj^{ik}_{ab} \vj^j_{ba} f_{ab} }{ \omega_1 +i\eta -\epsilon_{ba}} \\
             &+\sum_{a,b}\frac{1}{2}  \frac{\vj^{i}_{ab} \vj^{jk}_{ba} f_{ab} }{ \omega + 2i\eta -\epsilon_{ba}}\\
             &+\sum_{a,b,c} \frac{1}{2}\frac{\vj^{i}_{ab}}{ \omega + 2i\eta -\epsilon_{ba}} \left(   \frac{\vj^j_{bc}\vj^{k}_{ca}f_{ac} }{\omega_2 + i \eta -\epsilon_{ca} }    -    \frac{\vj^j_{ca}\vj^{k}_{bc}f_{cb} }{\omega_2 + i \eta -\epsilon_{bc} }  \right)\\
             &+\sum_{a,b,c}\frac{1}{2}\frac{\vj^{i}_{ab}}{ \omega + 2i\eta -\epsilon_{ba}} \left(  \frac{\vj^k_{bc}\vj^{j}_{ca}f_{ac} }{\omega_1 + i \eta -\epsilon_{ca} }    -    \frac{\vj^k_{ca}\vj^{j}_{bc}f_{cb} }{\omega_1 + i \eta -\epsilon_{bc} } \right)  \Biggr].
       \end{align*}
Let us consider the case of rectification responses $(\omega = 0,~\omega_1 =- \omega_2 =-\omega )$.
Since the resonant contribution breaks down the non-absorption condition, we drop the contributions including $\delta (\omega - \epsilon_{ab})$ arising from $\left( \omega + i\eta -\epsilon_{ab} \right)^{-1}$, i.e., $\kappa^{\rm a}_{ijk}(0;-\omega,\omega)$.
As a result, the rectification-response function is~\cite{de-Juan2020-ev,Sipe2000-sb,Watanabe2021-bt,Watanabe2022-hk}
        \begin{align}
        \kappa_{ijk}^\text{na} (0;-\omega,\omega)
             &= \frac{1}{2} \lim_{\bm{F}\to \bm{0}} \Biggl[ \sum_a \frac{1}{2}  \vj^{ijk}_{aa} f_a +\sum_{a,b}  \frac{\vj^{ij}_{ab} \vj^k_{ba} f_{ab} }{ \omega -\epsilon_{ba}} +\sum_{a,b}^{a\neq b}\frac{1}{2}  \frac{\vj^{i}_{ab} \vj^{jk}_{ba} f_{ab} }{ \epsilon_{ab}}\notag\\
             & +\frac{1}{2} \sum_{a,b} \partial_{F_i} \left( \frac{1}{\omega - \epsilon_{ba}} \right) X_{ab}^j X_{ba}^k f_{ab}  +\sum_{a,b,c}^{a\neq b} \frac{\vj^{i}_{ab}}{ \epsilon_{ab}} \left(   \frac{\vj^j_{bc}\vj^{k}_{ca}f_{ac} }{\omega -\epsilon_{ca} }    -    \frac{\vj^j_{ca}\vj^{k}_{bc}f_{cb} }{\omega -\epsilon_{bc} }  \right)  \Biggr] \notag \\
             &+ \left[ \left( \omega, j,k \right) \leftrightarrow \left( -\omega, k,j \right) \right] . \label{App_rectification_reactive_general}
       \end{align}
The energy and eigenstate are parametrized by the auxiliary fields $\bm{F}$~\cite{Watanabe2022-hk}, and hence the Hellmann-Feynman relation is obtained as
                \begin{equation}
                \vj_{ab}^i = \epsilon_{ab} \Braket{\partial_{F_i} a | b},
                \label{Hellmann_Feynman} 
                \end{equation}
between the states $(a,b)$ with different eigenvalues.
Then, the fifth term of Eq.~\eqref{App_rectification_reactive_general} including three eigenstates $(a,b,c)$ is transformed into
                \begin{equation}
                        \sum_{a,b,c}^{a\neq b} \frac{\vj^{i}_{ab}}{ \epsilon_{ab}} \left(   \frac{\vj^j_{bc}\vj^{k}_{ca}f_{ac} }{\omega -\epsilon_{ca} }    -    \frac{\vj^j_{ca}\vj^{k}_{bc}f_{cb} }{\omega -\epsilon_{bc} }  \right) = \sum_{a\neq b}\left( D_{F_i} \vj_{ab}^j - \vj_{ab}^{ij} \right) \vj_{ba}^k \frac{f_{ab}}{\omega -\epsilon_{ba}},
                \end{equation}
where we defined $D_{F_i}X_{ab}^j=\braket{\partial_{F_i}a|(1-\ket{a}\bra{a})X^j|b}+\braket{a|X^{ij}|b}+\braket{a|X_j(1-\ket{b}\bra{b})|\partial_{F_i}b}$.
One can notice that this term is partially canceled out by the second term of Eq.~\eqref{App_rectification_reactive_general}. 
Similarly, the third term is recast as
                \begin{equation}
                        \sum_{a,b}^{a\neq b}\frac{1}{2}  \frac{\vj^{i}_{ab} \vj^{jk}_{ba} f_{ab} }{ \epsilon_{ab}} = \frac{1}{2}\sum_a \left(  \partial_{F_i} \vj_{aa}^{jk} -\vj_{aa}^{ijk} \right) f_a.
                \end{equation}
The component including $\vj_{aa}^{ijk}$ is canceled out by the first term of Eq.~\eqref{App_rectification_reactive_general}.
Then, resuming all the terms and using $D_{F_i}X^{j}_{ab}X_{ba}^k+X^j_{ab}D_{F_i}X_{ba}^k=\partial_{F_i}(X^j_{ab}X^k_{ba})$, we arrive at the final expression 
                \begin{align}
                \kappa^\text{na}_{ijk} (0;-\omega,\omega)
                &= \frac{1}{2} \lim_{\bm{F}\to \bm{0}} \Biggl[ \sum_a \partial_{F_i}  \vj^{jk}_{aa} f_a - \sum_{a \neq  b} \vj_{ba}^j \vj_{ab}^k f_{ab} \partial_{F_i} \left( \frac{1}{\omega+\epsilon_{ba}} \right) - \partial_{F_i} \left( \vj_{ba}^j \vj_{ab}^k \right) \frac{f_{ab}}{\omega+ \epsilon_{ba}} \Biggr],\\
                &= \frac{1}{2} \lim_{\bm{F}\to \bm{0}} \Biggl[ \partial_{F_i} \left\{ \sum_a   \vj^{jk}_{aa} f_a  - \sum_{a \neq  b} \vj_{ba}^j \vj_{ab}^k \frac{f_{ab} }{\omega+\epsilon_{ba}} \right\} - \sum_a   \vj^{jk}_{aa} \partial_{F_i} f_a  -  \sum_{a \neq  b} \vj_{ab}^j \vj_{ba}^k  \frac{1}{\omega+\epsilon_{ab}} \partial_{F_i} f_{ab} \Biggr].
                \end{align}
The non-absorptive Pockels-response function of Eq.~\eqref{off-resonant_Pockels} is derived in a similar manner.

\clearpage

\end{document}